\pdfoutput=1
\documentclass[a4paper,american,floatfix,pdftex,superscriptaddress,twoside,%
aip,jcp,%
citeautoscript,
longbibliography,%
reprint,
]{revtex4-1}%
\usepackage{amsfonts,amsmath,amssymb}
\usepackage{bm}
\usepackage[T1]{fontenc}
\usepackage{lmodern}
\usepackage{graphicx}%
\usepackage[utf8]{inputenc}
\usepackage{microtype}
\usepackage[obeyFinal,textsize=footnotesize]{todonotes}
\usepackage{xspace}
\usepackage{hyperref, hypernat}
\usepackage[ams,todos]{CMImacros}

\newcommand{\cfeldesy}{\affiliation{Center for Free-Electron Laser Science, Deutsches
      Elektronen-Synchrotron DESY, Notkestraße 85, 22607 Hamburg, Germany}}%
\newcommand{\uhhcui}{\affiliation{Center for Ultrafast Imaging, Universität Hamburg, Luruper
      Chaussee 149, 22761 Hamburg, Germany}}%
\newcommand{\uhhphys}{\affiliation{Department of Physics, Universität Hamburg, Luruper Chaussee 149,
      22761 Hamburg, Germany}}%
\newcommand{\mbi}{\affiliation{Max-Born-Institut für nichtlineare Optik und Kurzzeitspektroskopie,
      Max-Born-Strasse 2a, 12489 Berlin, Germany}}%
\newcommand{\aachem}{\affiliation{Department of Chemistry, Aarhus University, Langelandsgade 140,
      DK-8000 Aarhus C, Denmark}}%
\newcommand{\aremail}{\email{rouzee@mbi-berlin.de}}%
\newcommand{\jkemail}{\email{jochen.kuepper@cfel.de}}%
\newcommand{\cmiweb}{\homepage{https://www.controlled-molecule-imaging.org}}%

\begin{document}
\title{Atomic-resolution imaging of carbonyl sulfide by laser-induced electron diffraction}
\author{Evangelos T. Karamatskos}\cfeldesy\uhhphys\uhhcui%
\author{Gildas Goldsztejn}\mbi%
\author{Sebastian Raabe}\mbi%
\author{Philipp Stammer}\mbi%
\author{Terry~Mullins}\cfeldesy%
\author{Andrea~Trabattoni}\cfeldesy%
\author{Rasmus R. Johansen}\aachem%
\author{Henrik Stapelfeldt}\aachem%
\author{Sebastian~Trippel}\cfeldesy\uhhcui%
\author{Marc J.~J.\ Vrakking}\mbi%
\author{Jochen~Küpper}\jkemail\cmiweb\cfeldesy\uhhphys\uhhcui%
\author{Arnaud~Rouz\'{e}e}\aremail\mbi%
\date{\today}
\begin{abstract}
   Measurements on the strong-field ionization of carbonyl sulfide molecules by short, intense,
   2~\um wavelength laser pulses are presented from experiments where angle-resolved photoelectron
   distributions were recorded with a high-energy velocity map imaging spectrometer, designed to
   reach a maximum kinetic energy of 500~eV. The laser-field-free elastic-scattering cross section
   of carbonyl sulfide was extracted from the measurements and is found in good agreement with
   previous experiments, performed using conventional electron diffraction. By comparing our
   measurements to the results of calculations, based on the quantitative rescattering theory (QRS),
   the bond lengths and molecular geometry were extracted from the experimental differential cross
   sections to a precision better than $\pm5$~pm and in agreement with the known values.
\end{abstract}
\maketitle

\section{Introduction}
\label{sec:intro}
Probing the structure of small to medium-size molecules in the gas phase is a formidable challenge.
Femtosecond x-ray diffractive-imaging experiments have been recently demonstrated at free-electron
lasers (FELs)~\cite{Kuepper:PRL112:083002, Glownia:PRL117:153003} and first electron-diffraction
experiments using relativistic electron beams have been performed~\cite{Weathersby:RSI86:073702,
   Yang:PRL117:153002}. Alternatively, new laser-based approaches are being developed that make use
of strong-field ionization with intense and ultrashort laser pulses. In a strong laser field, an
electron can tunnel ionize near the peak of the oscillatory laser field and is then accelerated in
the field. Depending on the ionization time within the laser period, the electron can be driven back
by the laser field and elastically scatter from its parent ion~\cite{Corkum:PRL71:1994}. The
scattered electron is accelerated again in the laser field, reaching a very high kinetic energy.
This process is responsible for the appearance of a recollision plateau in the photoelectron
momentum distribution. Extracting the differential scattering cross section (DCS) of the molecule
from the angular distribution of these high-energy electrons allows to derive the molecular
structure~\cite{Spanner:JPB37:L243, Blaga:Nature483:194}.

Laser-induced electron diffraction (LIED) has been first applied in
atoms~\cite{Okunishi:PRL100:143001, Xu:PRL109:233002}, in diatomic
molecules~\cite{Blaga:Nature483:194, Xu:NatComm5:4635} and more recently in polyatomic molecular
systems, such as acetylene~\cite{Pullen:NatComm6:7262}, ethylene~\cite{Ito:PRA96:053414}, and
benzene~\cite{Ito:SD3:034303}. To extract structural information from a LIED experiment, the
returning electron wavepacket should have a de~Broglie wavelength comparable with the bond lengths
that occur in the molecule. This is typically achieved using a mid-infrared laser field, as the
ponderomotive energy $\Up{}/\text{eV}=9.33\,I/(10^{14}~\Wpcmcm)\,\lambda^2/\um^2$ scales
quadratically with the laser wavelength.

While coupled rotational-electronic wavepacket dynamics in NO~\cite{Walt:NatComm8:15651} and
ultrafast bond breaking in acetylene dications in the presence of the strong ionizing laser
field~\cite{Wolter:Science354:308} have been recently characterized by LIED, the suitability of this
technique for retrieving transient molecular structures following photoexcitation has yet to be
demonstrated. In this context, carbonyl sulfide (OCS) is a particularly interesting system.
Photoexcitation of ground-state OCS ($X^1\Sigma$) in the (220–250~nm) UV wavelength range has been
intensively investigated~\cite{Brouard:JCP127:084304,Suzuki:JCP109:5778,Sivakumar:JPC89:3609} and is
known to be dominated by a transition to the $A^{'}$ state, leading to fragmentation of the
molecules predominantly into
$\text{CO}(X\,^{1}\Sigma^{+})+\text{S}(^{1}D_{2})$~\cite{Brouard:JCP127:084304}. Non-adiabatic
coupling along the bending coordinate is responsible for the formation of low-speed
$\text{S}(^{1}D_{2})$ fragments~\cite{Suzuki:JCP109:5778} and the production of rotationally excited
CO fragments~\cite{Sivakumar:JPC89:3609}. Therefore, OCS can serve as a benchmark to test the
suitability of the LIED method for recording molecular movies of molecular dynamics, in this case
imaging of molecular dissociation involving bending motion.

Interestingly, strong-field ionization of OCS, performed with 800~nm radiation and intensities above
$10^{15}~\Wpcmcm$, showed evidence for a bending deformation of the molecule during Coulomb
explosion~\cite{Sanderson:PRA65:043403}. This suggests that the molecular geometry can substantially
change both during and after ionization, with a large impact on the retrieval of the molecular
structure in a LIED experiment. Therefore, it is important to test the ability of the LIED technique
to image the structure of the OCS molecule in its equilibrium geometry before performing any
dynamical investigation.

Here, we present an LIED measurement on OCS molecules, photoionized by strong 2~\um wavelength laser
pulses. Different from previous investigations~\cite{Blaga:Nature483:194, Pullen:NatComm6:7262,
   Wolter:Science354:308}, the experiment was performed using a velocity map imaging
spectrometer~\cite{Eppink:RSI68:3477} that allowed the detection of all electrons with kinetic
energies up to 500~eV. Our new experimental apparatus was benchmarked by strong-field ionization
experiments on argon and krypton. DCSs extracted from measured photoelectron angular distributions
for a recolliding electron energy of 100~eV were compared to results from partial-wave calculations
for scattering of electrons by atoms performed using the Elsepa package~\cite{Salvat:CPC165:157},
and agree very well. The same procedure was applied to the experimental data for OCS. Fitting the
calculated DCS to the experimental one with the O-C and C-S bond lengths and the
$\angle\text{(O-C-S)}$ bending angle as adjustable parameters, we were able to confirm the linear
structure and were able to extract the internuclear distances of the molecule to a precision better
than 5~pm. Our experimentally determined values agree very well with reported bond lengths of the
OCS molecule~\cite{Dakin:PR71:640} and suggest that accurate structures can be retrieved for OCS
using the LIED technique.

\section{Experimental setup}
\label{sec:setup}
\begin{figure}
   \includegraphics[width=\linewidth]{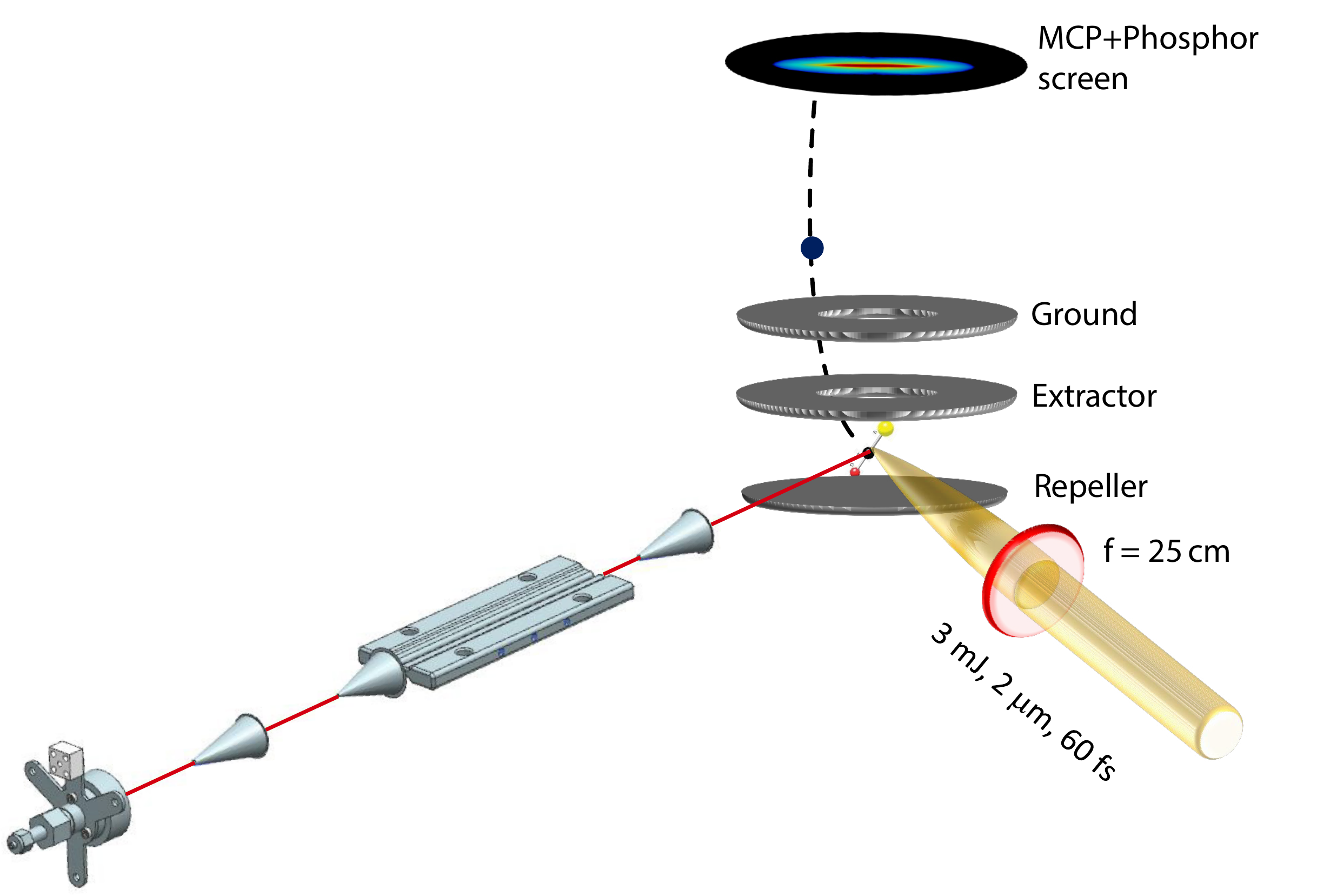}
   \caption{Sketch of the experimental setup. 2~\um laser pulses are focused into a beam of
      state-selected OCS molecules at the center of a velocity map imaging spectrometer. The
      molecular beam is formed by supersonic expansion of a dilute mixture of OCS in helium. A
      deflector and a skimmer are used to spatially separate the OCS molecules from the helium
      carrier gas. The photoelectrons are projected onto a multichannel-plate/phosphor-screen
      assembly and the resulting 2D electron momentum distribution is recorded with a CCD camera.}
   \label{fig:setup}
\end{figure}
The experiments were performed using 2~\um radiation pulses, obtained from an optical parametric
amplifier pumped by 800~nm laser pulses from a commercial amplifier system, delivering 30~mJ, 38~fs
(full width at half maximum, FWHM) pulses at a repetition rate of 1~kHz. The linearly polarized
2~\um laser pulses were focused into a beam of nearly pure ground-state OCS molecules at the center
of a high-energy velocity map imaging spectrometer (VMI) using a CaF$_2$ lens with a 25~cm focal
length, see \autoref{fig:setup}. The laser polarization axis was aligned in the plane of the
detector, corresponding to the vertical axis in all images presented, \emph{vide infra}.

The molecular beam was formed by supersonic expansion of a mixture of OCS in helium with a mixing
ratio 1/2000 and a constant pressure of 90~bar, using an Even-Lavie valve running at a repetition
rate of 250~Hz. In the experiments performed on atoms, a pure sample of either argon or krypton was
expanded into vacuum, where the stagnation pressure was limited to 1~bar to avoid cluster formation.
An electrostatic deflector~\cite{Kienitz:JCP147:024304} was operated at $\pm13.5$~kV to spatially
separate OCS molecules from the helium carrier gas~\cite{Chang:IRPC34:557}. We note that this device
is ideally suited for quantum-state selection and the preparation of structurally pure samples, even
of complex molecules~\cite{Chang:IRPC34:557, Teschmit:ACIE57:13775, Trippel:RSI89:096110}. Here, the
deflection provided the sample of OCS molecules in their rotational
ground-state~\cite{Nielsen:PCCP13:18971, Karamatskos:arXiv1807:01034}.

Photoelectrons from strong-field ionization (SFI) by the 2~\um laser pulses were accelerated into a
10 cm long field-free flight tube before being detected on a 77~mm diameter dual
microchannel-plate/phosphor-screen assembly. The projected 2D electron momentum distributions were
recorded using a CCD camera and inverted using an Abel-inversion procedure based on the BASEX
algorithm~\cite{Dribinski:RSI73:2634} in order to yield 3D electron momentum distributions.

\section{Results and discussion}
\label{sec:results}
\begin{figure}
   \includegraphics[width=\linewidth]{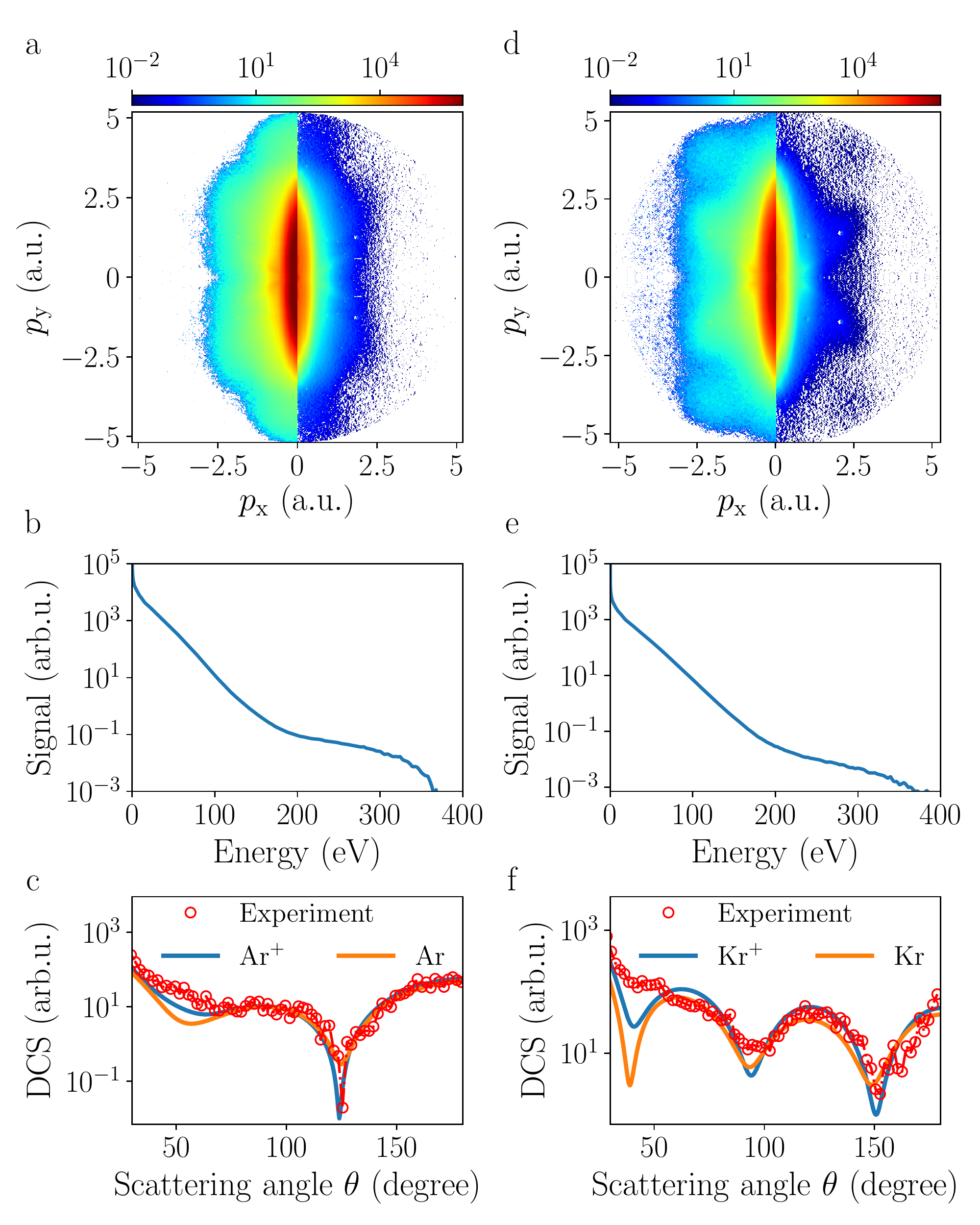}
   \caption{Projected 2D electron momentum distribution (left) and slice through the 3D electron
      momentum distribution obtained after Abel inversion (right) recorded for (a) argon and (d)
      krypton, ionized by intense 2~\um laser pulses. (b,~e) Corresponding photoelectron kinetic
      energy spectra. (c,~f) Field-free differential cross sections extracted from the
      angle-resolved photoelectron kinetic energy spectra (red dots) for electrons with a return
      energy of $\ordsim100$~eV. The DCS calculated using the Elsepa package is shown, for
      comparison, for a neutral atom (orange line) and a singly charged ion (blue line); see text
      for details.}
   \label{fig:LIED:atom}
\end{figure}
In order to validate our experimental methodology, experiments were first performed for pure samples
of rare-gas atoms. Projected 2D electron momentum distributions recorded in argon and krypton and
corresponding slices through their 3D momentum distributions are presented in
\autoref[a,~d]{fig:LIED:atom}. The laser intensity was $\ordsim1\times10^{14}~\Wpcmcm$, which
corresponds to a Keldysh parameter of $\gamma=\sqrt{I_\text{p}/2\Up{}}\approx0.38$, with the
ponderomotive energy $\Up{}=37.3$~eV, and $I_\text{p}=11.2$~eV the ionization potential, indicating
that the experiment was performed deep into the tunneling regime. The angle-resolved photoelectron
spectra were averaged over $10^{6}$ laser shots. To account for rest gas, an image obtained without
atomic beam was subtracted from the 2D electron momentum distributions prior to Abel inversion.

In a classical picture of the strong-field ionization~\cite{Corkum:PRL71:1994}, electrons that have
experienced a single recollision with the parent ion can reach a maximum kinetic energy of \Up{10},
whereas electrons that do not further interact with the parent ion -- commonly called ``direct
electrons'' -- can have a maximum kinetic energy of \Up{2}. In our measurement, the direct electron
yield is five to six orders of magnitude larger than the contribution from rescattered electrons,
see \autoref{fig:LIED:atom}. For argon and krypton, the photoelectron spectra observed
experimentally extend to a kinetic energy close to 400~eV.

Field-free DCSs of argon and krypton were extracted from our measurements following a procedure
given by the quantitative rescattering theory~\cite{Chen:PRA79:033409, Xu:PRA82:033403}. The
high-energy rescattered photoelectron momentum distribution $D(k,\theta)$ is expressed as the
product of the momentum distribution $W(k_r)$ of the returning electron (with $k_r$ the momentum at
the instant of recollision) and the DCS $\sigma(k_r,\theta_r)$, with $\theta_r$ the scattering
angle. The relationship between the measured electron momentum $k$ and $k_r$ is obtained by
considering that the scattered electrons gain an additional momentum after the recollision, which is
given by the vector potential $-A(t_r)$ at the time of recollision $t_r$:
\begin{eqnarray}
  k_y &=&k\cos\theta=-A(t_r) + k_r\cos\theta_r \\
  k_x &=& k\sin\theta=k_r\sin\theta_r
\end{eqnarray}
with $y$ defined as the laser polarization axis. According to the classical equations of motion and
neglecting the effect of the Coulomb potential on the electron trajectories, the maximum recollision
electron momentum satisfies $k_r=1.26A_0$, with $A_0$ the magnitude of the vector potential,
corresponding to a maximum kinetic energy of $\Up{\ordsim3.17}$, \ie, $\ordsim118$~eV for a
wavelength of 2~\um and an intensity of $\ordsim1\times10^{14}~\Wpcmcm$. The DCS
$\sigma(k_r,\theta_r)$ for the highest recollision energy can, therefore, be extracted from the
photoelectron angular distribution (PAD) by measuring the photoelectron yield on a circle with
radius $k_r=1.26A_0$~\cite{Okunishi:PRL100:143001} and centered at $(k_{x},k_{y})=(0,\pm A(t_r))$.
We note that this procedure yields the DCS weighted by the ionization yield.

\autoref[c,~f]{fig:LIED:atom} shows the field-free DCS extracted for argon and krypton using this
method. The results were obtained using an integration range of $\Delta{}k_r\approx 0.05k_r$ and an
angular integration width of $\Delta\theta=\degree{1}$. For krypton two pronounced minima at
scattering angles of \degree{94} and \degree{151} are clearly observed. For argon, the DCS presents
a strong dip near \degree{124} and a broad minimum near \degree{60}. These results are in very good
agreement with previous LIED experiments~\cite{Xu:PRL109:233002} as well as with conventional
electron-scattering experiments using an external electron source~\cite{Fon:JPB16:307,
   Fon:JPB17:3279}. The DCS for both atomic targets compare also very well with theoretical
calculations for field-free electron--atom collisions obtained using the Elsepa
package~\cite{Salvat:CPC165:157}, shown as orange and blue curves in \autoref[c,~f]{fig:LIED:atom}.
In these calculations, the nuclear charge distribution was approximated by a point charge and the
electron charge density of the atomic cation was evaluated from self-consistent Dirac-Fock
calculations. Exchange and correlation-polarization potentials were neglected. The simulations were
performed considering both, a neutral and an ionic, atomic target. We note that to achieve the best
agreement with the experimental DCS, the magnitude of the vector potential and, therefore, the laser
intensity used to extract the DCS from the PAD was fitted. Best agreement was found for intensities
of $9.1\times10^{13}~\Wpcmcm$ and $8.3\times10^{13}~\Wpcmcm$, with corresponding return electron
kinetic energies of 98 and 107~eV, for argon and krypton, respectively. These values are in close
agreement with the estimated intensity based on the laser-pulse parameters used in these
experiments. We attribute the difference observed between the two atomic targets to a small
variation of the pulse energy between the two measurements. The comparison between the
experimentally retrieved DCS and the simulated DCS, obtained for a neutral and an ionic atomic
target shown in \autoref[c,~f]{fig:LIED:atom}, reveals that a better agreement is found when
considering that the returning electron interacts with a singly charged atomic ion.

\begin{figure}
   \includegraphics[width=\linewidth]{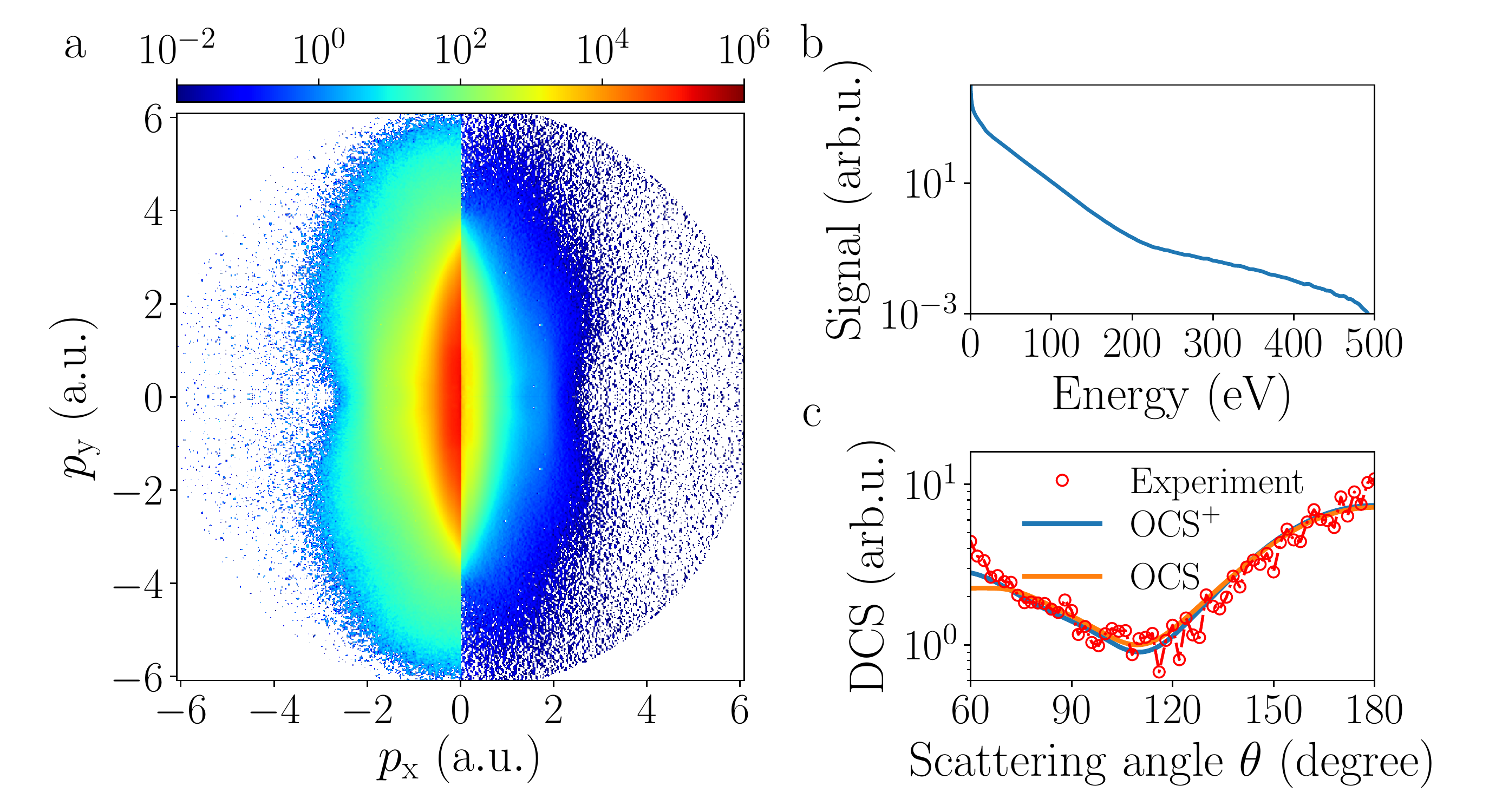}
   \caption{(a) Projected 2D electron momentum distribution (left) and slice through the 3D electron
      momentum distribution obtained after Abel inversion (right) recorded in OCS ionized by an
      intense 2~\um laser pulse with a laser intensity of $\ordsim1.3\times10^{14}~\Wpcmcm$. (b)
      Corresponding photoelectron kinetic energy spectra. (c) Field-free differential cross section
      extracted from the angle-resolved photoelectron kinetic energy spectra (dotted red line) for
      electrons with a return energy of $\ordsim100$~eV (c). The DCS calculated using the molecular
      Elsepa package that best fits the measurement is shown as well for neutral OCS (orange line)
      and for singly charged OCS$^{+}$, where the scattering amplitude of neutral sulfur was
      replaced by the corresponding ionic scattering amplitude (blue line).}
   \label{fig:LIED:OCS}
\end{figure}
Subsequently, we recorded the PAD resulting from SFI of OCS. Note that no laser alignment was used
in the experiment, and hence the OCS molecules were randomly oriented prior to their interaction
with the laser. The laser intensity was adjusted to observe only the parent molecular ion in an ion
time-of-flight measurement (fragmentation $<1\%$) in order to minimize the influence of multiple
ionization channels. The 2D momentum distribution recorded for OCS and its corresponding
photoelectron kinetic-energy spectrum are shown in \autoref[a,~b]{fig:LIED:OCS}. The kinetic energy
spectrum extends to 480~eV, suggesting an intensity $\ordsim1.3\times10^{14}~\Wpcmcm$, i.e.,
slightly higher than in the measurements for argon and krypton.

Similarly to the atomic case, the field-free DCS was retrieved from the PAD for a return electron
energy of 100~eV and is shown in \autoref[c]{fig:LIED:OCS}. A broad minimum near \degree{110} is
observed, similar to previously reported electron-scattering experiments with 100~eV kinetic energy
projectiles ~\cite{Mirai:JCP5:054302, Michelin:JPB33:3293}. The minimum observed at \degree{110} is
known to be dominated by the atomic form factor of the sulfur atom, smeared out by the molecular
structure~\cite{Mirai:JCP5:054302}. To extract the internuclear distances of OCS from our
measurement, we applied a procedure that was first introduced in
reference~\onlinecite{Blaga:Nature483:194} to retrieve the internuclear distance of diatomic
molecules from LIED measurements. For a fixed-in-space molecule, oriented at Euler angles $\Omega_L$
with respect to the laser polarization axis $y$, the PAD is written as:

\begin{equation}
   D(k,\theta,\Omega_L)=W(k_r)N(\Omega_L)\sigma(k_r,\theta_r,\Omega_L),
\end{equation}
with $N(\Omega_L)$ the angle-dependent ionization probability. For an isotropic molecular sample the
measured signal is then given by:
\begin{equation}
   I(k,\theta)=W(k_r)\int d\Omega_LN(\Omega_L)\sigma(k_r,\theta_r,\Omega_L).
   \label{eqn:liedsignal}
\end{equation}

Recent studies~\cite{Schell:SciAdv4:eaap8148} have shown that the shape of molecular orbitals can
leave its imprint on the recollision probability. Moreover, in molecular ionization, multiple
orbitals can contribute to ionization~\cite{Krecinic:PRA98:041401,Trabattoni:cutoff:inprep}. For
OCS, the HOMO (IP=11.2~eV) and HOMO-1 (15.1~eV) orbitals are separated by $\approx$4~eV and the
contribution of the HOMO-1 orbital to the ionization dynamics is expected to be negligible. Since
randomly oriented molecules were used in the experiment, we assume that the influence of the shape
of the molecular orbital from which the electron is emitted is washed out during the propagation of
the electron wavepacket in the laser field. Using this assumption, the field-free DCS
in~\eqref{eqn:liedsignal} can be approximated by an independent-atom model (IAM) and expressed as:
\begin{equation}
   \sigma(k_r,\theta_r,\Omega_L)=\sum_{i,j}f_i(\theta_r)f^*_j(\theta_r)e^{i\vec{q}\cdot\vec{R}_{ij}},
\end{equation}
with the momentum transfer $q=2k_r\sin(\theta_r/2)$, the internuclear distances $R_{ij}$ and the
scattering amplitude $f_{i}(\theta_r)$ for atom $i$. The returning electron interacts with the
molecular ion, which we modeled by a singly charged sulfur atom and neutral carbon and oxygen atoms.
This is well justified as the removal of an electron from the HOMO of OCS is expected to lead to a
molecular ion with a final charge mainly localized on the sulfur
atom~\cite{Holmegaard:NatPhys:2010}, see the appendix for further details.

Combining the IAM with the QRS yields the following expression that was used for the analysis of our
measurements:
\begin{multline}
   I(k,\theta)=W(k_r)\left(\sum_i\abs{f_i}^2\int N(\Omega_L)d\Omega_L \right. \\
   + \left. \sum_{i\neq j}f_if^*_j\int N(\Omega_L)e^{i\vec{q}\cdot\vec{R}_{ij}}d\Omega_L\right).
   \label{DCSth}
\end{multline}
The first term corresponds to an incoherent sum over the scattering amplitudes $I_\text{atom}$ of
the individual atoms whereas the second term corresponds to a molecular interference term. Following
the standard approach~\citep{Blaga:Nature483:194}, we define the molecular contrast factor (MCF)
$\gamma_{\text{MCF}}$ as:
\begin{equation}
   \gamma_{\text{MCF}} = \frac{I-I_\text{atom}}{I_\text{atom}}
   = \frac{\sum_{i\neq j}f_if^*_j\int N(\Omega_L)e^{i\vec{q}\cdot\vec{R}_{ij}}d\Omega_L}{\sum_i\vert f_i\vert^2\int N(\Omega_L)d\Omega_L}.
   \label{MCF}
\end{equation}
\begin{figure}
   \includegraphics[width=0.8\linewidth]{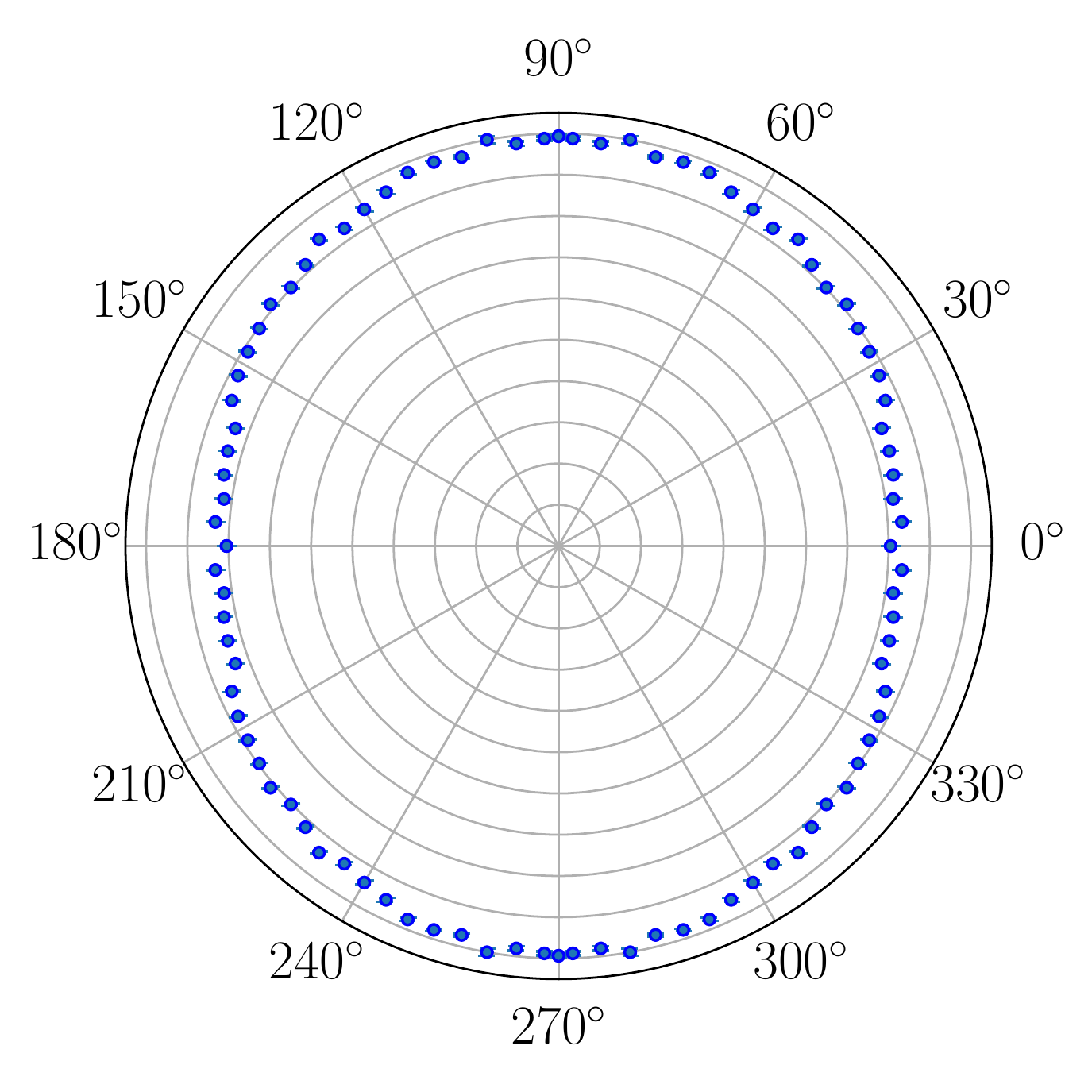}
   \caption{Measured ionization rate from strong-field ionization of OCS by a linearly polarized
      2~\um laser pulse with a laser intensity of $\ordsim1.3\times10^{14}~\Wpcmcm$ as a function of
      the angle between the internuclear axis and the laser polarization axis.}
   \label{fig:ADIP:OCS}
\end{figure}

\begin{figure}
   \includegraphics[width=\linewidth]{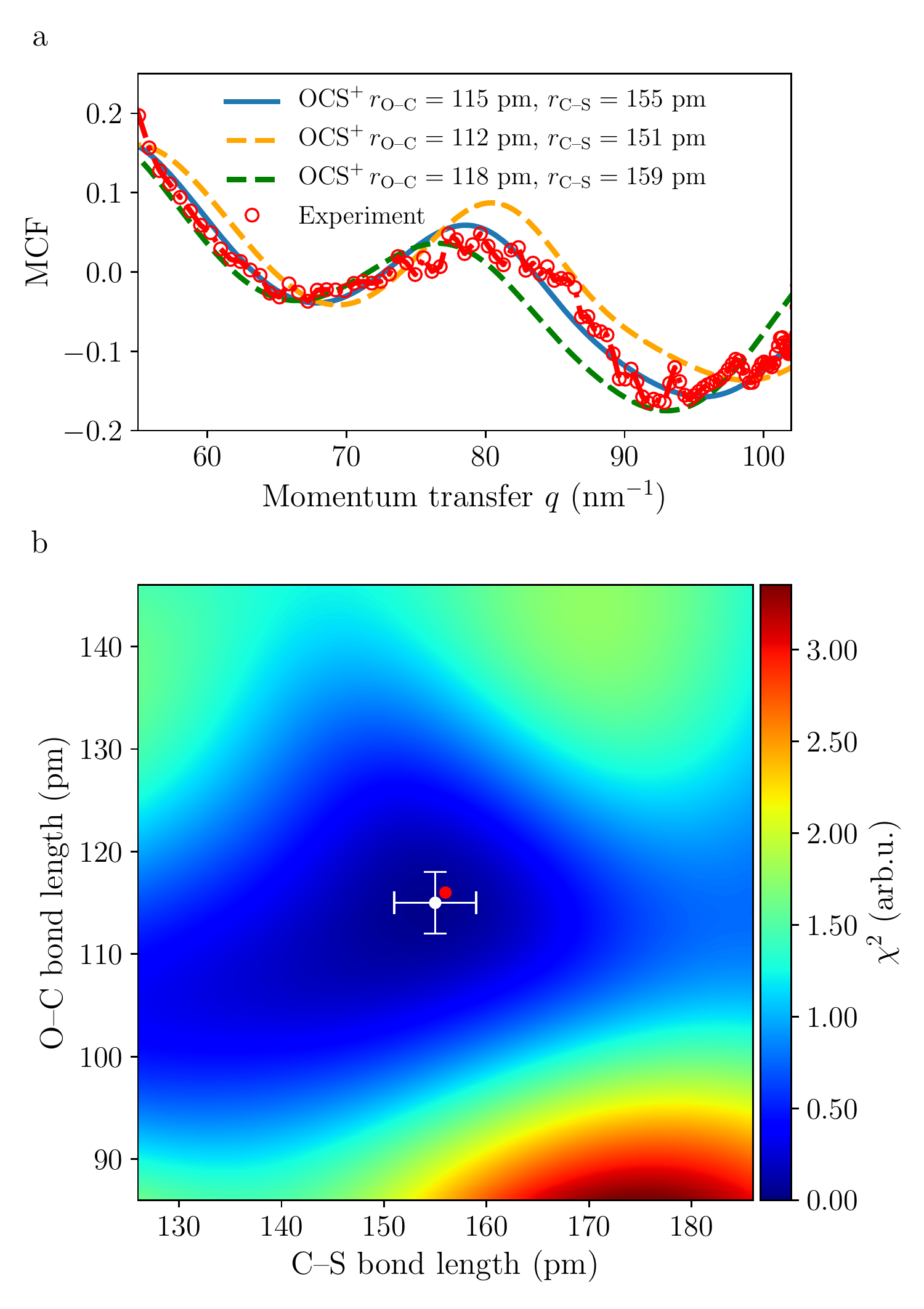}
   \caption{(a) Comparison between the MCF extracted from the experiment (red open circles) and the
      calculated molecular contrast factor obtained for the best fit of the bond lengths (blue
      line). Dashed lines depict the expected behavior for bond lengths changed by $\pm\sigma$
      (green and orange). (b) $\chi^2$ map as a function of the bond lengths considering a linear
      configuration of the molecule. The minimum (white dot) corresponds to the best fit and the
      crossed bars depict the $1\sigma$ error; this result agrees very well with reference values
      for the internuclear distances of OCS (red dot).}
   \label{fig:LIED:chi2}
\end{figure}

In order to extract bond lengths from our measurement, we have compared the MCF extracted
experimentally with simulations using expression~\eqref{MCF}. The neutral atomic scattering factors
were obtained using the Elsepa package~\cite{Salvat:CPC165:157}. To estimate the angle-dependent
ionization probability $N(\Omega_L)$ necessary to calculate the MCF, the following experiment was
performed. A sequence of two laser pulses, at a wavelength centered at 800~nm and with 255~fs pulse
duration, were used to strongly align the molecule prior to the 2~\um laser pulse, see reference
~\cite{Karamatskos:arXiv1807:01034} for details. The angle-dependent ionization probability was then
obtained experimentally by monitoring the ionization yield as a function of the angle between the
molecular axis and the ionizing laser polarization, see \autoref{fig:ADIP:OCS}, and then used to
calculate the MCF. Finally, the $R_\text{O--C}$ and $R_\text{C--S}$ bond distances were fitted in
order to minimize the variance between experiment and theory using the following expression for the
error:

\begin{align}
  \chi^2(\beta,R_{ij}) = (\gamma^{exp}_{\text{MCF}}-\gamma^{th}_{\text{MCF}})^2=\left(\frac{\beta I_{\text{exp}}-I_{\text{th}}}{I_{\text{atom}}}\right)^2,
\end{align}
with $\beta$ a normalization constant, $I_{\text{exp}}$ the DCS extracted from the measured
photoelectron spectrum and $I_{\text{th}}$ the DCS calculated using~\eqref{DCSth}. The result from
this procedure is shown in~\autoref{fig:LIED:chi2}. The best agreement is obtained for
\mbox{$R_\text{O--C}=115\pm3$~pm} and \mbox{$R_\text{C--S}=155\pm4$~pm}. Even for the relatively low
return electron energy of 100~eV, a precision of $\pm4$~pm is reached. These values are in very
close agreement with the known values \mbox{$R_\text{O--C}=116\pm2$~pm} and
\mbox{$R_\text{C--S}=156\pm3$~pm} obtained by microwave absorption
spectroscopy~\cite{Dakin:PR71:640}, which are marked by the red dot in~\autoref{fig:LIED:chi2} b.

As previously mentioned, we cannot \emph{a priori} exclude a possible deformation of the molecule
following its ionization by the intense laser field. If this deformation takes place on a timescale
shorter than the duration between ionization and recollision, it would be observed in our experiment
as a deformed, bent or stretched OCS geometry upon recollision. However, we did not observe any
indication of stretching of the bond distances of the molecule, even when we performed an extended
analysis of our measurement using the overall O--S distance as an additional fitting parameter. Best
agreement was found for $R_\text{O--S}=270$~pm, with $R_\text{O--C}=114\pm4$~pm and
$R_\text{C--S}=155\pm5$~pm, \ie, for the linear configuration of the molecule. This suggests that in
our experiment, with intensity $\ordsim1\times10^{14}~\Wpcmcm$ and wavelength 2~\um, corresponding
to a laser period of 6.6~fs, the molecular structure remains essentially unchanged during the time
interval between ionization to recollision. In this context we note that recent \emph{ab-initio}
calculations~\cite{Bilalbegovic:EPJD49:43} for laser pulses centered at 790~nm and an intensity of
$1\times10^{15}~\Wpcmcm$ have shown that the atomic distances and bending angle
$\angle\text{(O-C-S)}$ are changing on a timescale longer than 10~fs, \ie, longer than the optical
period in our experiment.

\section{Conclusion}
\label{sec:conclusion}
We have recorded angle-resolved photoelectron spectra of argon, krypton and OCS, ionized by short
laser pulses at 2~\um, with a high-energy VMI. We extracted field-free differential electron
rescattering cross sections at 100~eV, which are in excellent agreement with calculated DCSs for
electron-atom and electron-molecule scattering. The geometry and bond distances of the OCS molecule
were extracted from our measurement with a precision better than $\pm5$~pm and in full agreement
with the known structure of ground-state OCS.

It remains an open question to what extent the LIED technique can be used to retrieve multiple bond
lengths and angles during molecular transformations, for instance following the photoexcitation of a
molecule. Further investigations combining pump-probe schemes and LIED are ongoing to explore the
possibility to use this technique to directly record a so-called ``molecular movie'' of these
motions, in which the evolving structure is measured with femtosecond and picometer precision while
the molecule is ``in action''.

\section{Acknowledgements}
This work has been supported by the Deutsche Forschungsgemeinschaft (DFG) through the priority
program ``Quantum Dynamics in Tailored Intense Fields'' (QUTIF, SPP1840, AR 4577/4, KU 1527/3) and
by the Clusters of Excellence ``Center for Ultrafast Imaging'' (CUI, EXC 1074, ID 194651731) and
``Advanced Imaging of Matter'' (AIM, EXC 2056, ID 390715994) of the Deutsche Forschungsgemeinschaft,
by the Helmholtz Gemeinschaft through the ``Impuls- und Vernetzungsfond'', and by the European
Union's Horizon 2020 research and innovation programme under the Marie Sklodowska-Curie grant
agreement No 674960 (ASPIRE). A.T. gratefully acknowledges a fellowship of the Alexander von
Humboldt Foundation.

\appendix
\section{Charge distribution in the independent-atom model}
Structure retrieval in LIED experiments is typically achieved by employing the quantitative
rescattering theory (QRS)~\cite{Chen:PRA79:033409, Xu:PRA82:033403} combined with the
independent-atom model (IAM)~\cite{Hargittai:GED:1988,Lin:JPB43:122001}. Generally, the IAM is not
well suited to incorporate scattering from singly charged cations, as in this model a molecule is
described as a collection of independent atoms as the scattering centers for the incoming electron
flux. However, the hole charge density is generally delocalized within the molecule, which cannot be
described within the independent-atom model, where the substitution of neutral scattering amplitudes
by the corresponding ones for singly charged ions leads to a strong localization of the hole charge
density on one atomic site.

For OCS, it was shown that ionization from the HOMO of OCS leads to a molecular cation with
$\ordsim85~\%$ of the hole charge density localized at the S atom~\cite{Holmegaard:NatPhys:2010}.
Therefore, in the main article, the IAM model was applied by replacing the scattering amplitude of
neutral sulfur by the corresponding ionic one. Here, for comparison, we provide the results obtained
for a model in which OCS is either neutral or a singly charged ion with a final charge localized on
the carbon or the oxygen sites. Except for neutral OCS, the other cases did not allow to retrieve
the correct equilibrium geometry of OCS, confirming that the hole charge density of singly charged
OCS is mostly localized at the sulfur site.

\subsection{Neutral OCS}
\begin{figure}[b]
\centering
   \includegraphics[width=\linewidth]{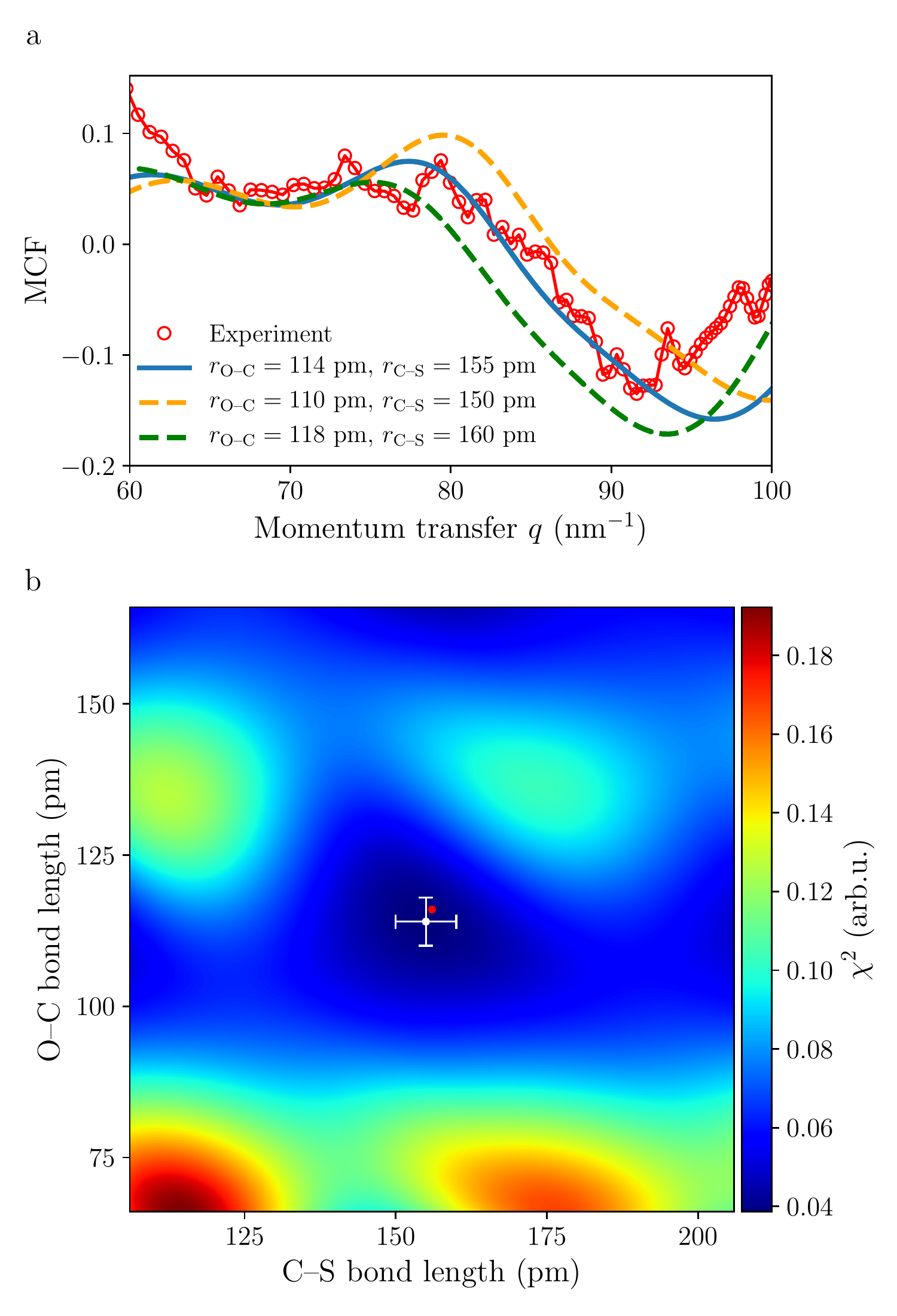}
   \caption{Same as~\autoref{fig:LIED:chi2} for the IAM model with neutral-atom scattering
      amplitudes for all atoms.}
   \label{fig1_SI}
\end{figure}
\autoref{fig1_SI} shows the results of the fitting procedure for neutral OCS, including only atomic
scattering amplitudes of neutral oxygen, carbon, and sulfur. The experimental MCF fits the
simulations only in the range $q=65\ldots90~\text{nm}^{-1}$ and a deviation for smaller and larger
momentum transfer is observed. The fits were thus carried out only in this range of momentum
transfer. Then, the best fit yielded values for the bond lengths of $\text{O--C}=114\pm4~\text{pm}$
and $\text{C--S}=155\pm5~\text{pm}$ with $\chi^{2}=0.0386$. While this provided bond lengths close
to the known values~\cite{Dakin:PR71:640}, with this model the MCF is not well reproduced for low
and large momentum transfers, whereas the cation model presented in the main text provides a robust
overall description of the experimental data.

\subsection{OCS$^{+}$ with the charge localized on O or C}
The same procedure was applied for a molecular cation with the charge localized on the oxygen atom,
see~\autoref[a,~b]{fig2_SI}, or on the carbon atom, see \autoref[c,~d]{fig2_SI}. The best fits
yielded values for the bond lengths of $\text{O--C}=130\pm5~\text{pm}$ and
$\text{C--S}=158\pm5~\text{pm}$ for a charge localized on the oxygen atom with $\chi^{2}=0.1161$;
and $\text{O--C}=43\pm5~\text{pm}$ and $\text{C--S}=124\pm 5~\text{pm}$ for a charge localized on
the carbon atom with $\chi^{2}=0.0595$. In these two cases, a large deviation from the known bond
distances of OCS is retrieved from the fit, highlighting the relevance of an appropriate hole charge
distribution in the analysis of LIED data. The best fit was obtained considering electron scattering
from a molecular cation with a final charge localized on the sulfur atom, see
\autoref{fig:LIED:chi2}, yielding bond lengths of $\text{O--C}=115\pm3~\text{pm}$ and
$\text{C--S}=155\pm5~\text{pm}$ with $\chi^{2}=0.0418$.
\begin{figure}[t]
   \includegraphics[width=\linewidth]{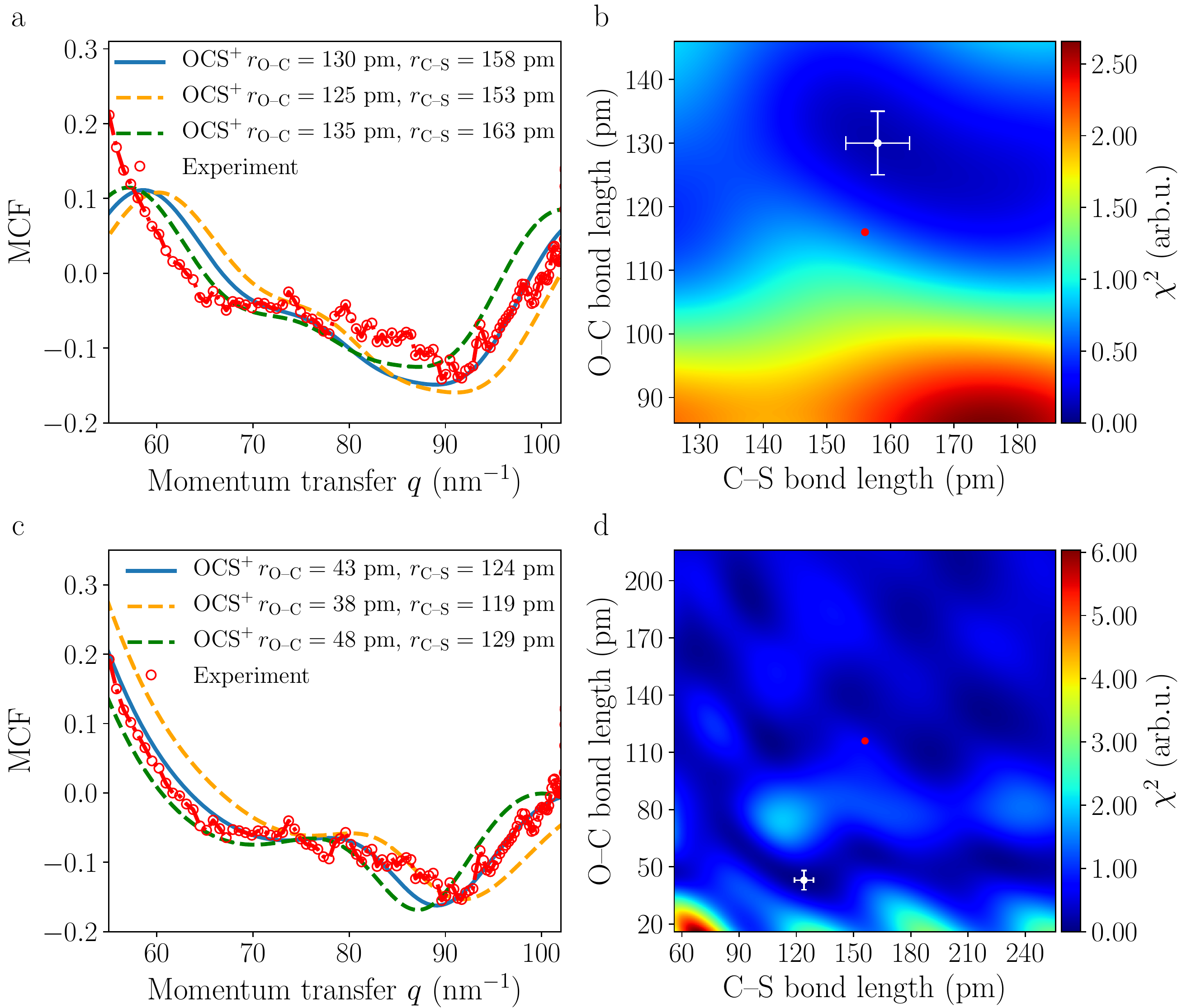}
   \caption{\textbf{a,~b} Same as~\autoref{fig:LIED:chi2}, considering electron scattering from a
      molecular cation with the charge localized on the oxygen atom. \textbf{c,~d} Same
      as~\autoref{fig:LIED:chi2}, considering electron scattering from a molecular cation with the
      charge localized on the carbon atom.}
   \label{fig2_SI}
\end{figure}

\bibliography{OCS-isotropic-LIED}%

\begin{thebibliography}{42}%
\makeatletter
\providecommand \@ifxundefined [1]{%
 \@ifx{#1\undefined}
}%
\providecommand \@ifnum [1]{%
 \ifnum #1\expandafter \@firstoftwo
 \else \expandafter \@secondoftwo
 \fi
}%
\providecommand \@ifx [1]{%
 \ifx #1\expandafter \@firstoftwo
 \else \expandafter \@secondoftwo
 \fi
}%
\providecommand \natexlab [1]{#1}%
\providecommand \enquote  [1]{``#1''}%
\providecommand \bibnamefont  [1]{#1}%
\providecommand \bibfnamefont [1]{#1}%
\providecommand \citenamefont [1]{#1}%
\providecommand \href@noop [0]{\@secondoftwo}%
\providecommand \href [0]{\begingroup \@sanitize@url \@href}%
\providecommand \@href[1]{\@@startlink{#1}\@@href}%
\providecommand \@@href[1]{\endgroup#1\@@endlink}%
\providecommand \@sanitize@url [0]{\catcode `\\12\catcode `\$12\catcode
  `\&12\catcode `\#12\catcode `\^12\catcode `\_12\catcode `\%12\relax}%
\providecommand \@@startlink[1]{}%
\providecommand \@@endlink[0]{}%
\providecommand \url  [0]{\begingroup\@sanitize@url \@url }%
\providecommand \@url [1]{\endgroup\@href {#1}{\urlprefix }}%
\providecommand \urlprefix  [0]{URL }%
\providecommand \Eprint [0]{\href }%
\providecommand \doibase [0]{http://dx.doi.org/}%
\providecommand \selectlanguage [0]{\@gobble}%
\providecommand \bibinfo  [0]{\@secondoftwo}%
\providecommand \bibfield  [0]{\@secondoftwo}%
\providecommand \translation [1]{[#1]}%
\providecommand \BibitemOpen [0]{}%
\providecommand \bibitemStop [0]{}%
\providecommand \bibitemNoStop [0]{.\EOS\space}%
\providecommand \EOS [0]{\spacefactor3000\relax}%
\providecommand \BibitemShut  [1]{\csname bibitem#1\endcsname}%
\let\auto@bib@innerbib\@empty
\bibitem [{\citenamefont {K{\"u}pper}\ \emph {et~al.}(2014)\citenamefont
  {K{\"u}pper}, \citenamefont {Stern}, \citenamefont {Holmegaard},
  \citenamefont {Filsinger}, \citenamefont {Rouz\'{e}e}, \citenamefont
  {Rudenko}, \citenamefont {Johnsson}, \citenamefont {Martin}, \citenamefont
  {Adolph}, \citenamefont {Aquila}, \citenamefont {Bajt}, \citenamefont
  {Barty}, \citenamefont {Bostedt}, \citenamefont {Bozek}, \citenamefont
  {Caleman}, \citenamefont {Coffee}, \citenamefont {Coppola}, \citenamefont
  {Delmas}, \citenamefont {Epp}, \citenamefont {Erk}, \citenamefont {Foucar},
  \citenamefont {Gorkhover}, \citenamefont {Gumprecht}, \citenamefont
  {Hartmann}, \citenamefont {Hartmann}, \citenamefont {Hauser}, \citenamefont
  {Holl}, \citenamefont {H{\"o}mke}, \citenamefont {Kimmel}, \citenamefont
  {Krasniqi}, \citenamefont {K{\"u}hnel}, \citenamefont {Maurer}, \citenamefont
  {Messerschmidt}, \citenamefont {Moshammer}, \citenamefont {Reich},
  \citenamefont {Rudek}, \citenamefont {Santra}, \citenamefont {Schlichting},
  \citenamefont {Schmidt}, \citenamefont {Schorb}, \citenamefont {Schulz},
  \citenamefont {Soltau}, \citenamefont {Spence}, \citenamefont {Starodub},
  \citenamefont {Str{\"u}der}, \citenamefont {Th{\o}gersen}, \citenamefont
  {Vrakking}, \citenamefont {Weidenspointner}, \citenamefont {White},
  \citenamefont {Wunderer}, \citenamefont {Meijer}, \citenamefont {Ullrich},
  \citenamefont {Stapelfeldt}, \citenamefont {Rolles},\ and\ \citenamefont
  {Chapman}}]{Kuepper:PRL112:083002}%
  \BibitemOpen
  \bibfield  {author} {\bibinfo {author} {\bibfnamefont {J.}~\bibnamefont
  {K{\"u}pper}}, \bibinfo {author} {\bibfnamefont {S.}~\bibnamefont {Stern}},
  \bibinfo {author} {\bibfnamefont {L.}~\bibnamefont {Holmegaard}}, \bibinfo
  {author} {\bibfnamefont {F.}~\bibnamefont {Filsinger}}, \bibinfo {author}
  {\bibfnamefont {A.}~\bibnamefont {Rouz\'{e}e}}, \bibinfo {author}
  {\bibfnamefont {A.}~\bibnamefont {Rudenko}}, \bibinfo {author} {\bibfnamefont
  {P.}~\bibnamefont {Johnsson}}, \bibinfo {author} {\bibfnamefont {A.~V.}\
  \bibnamefont {Martin}}, \bibinfo {author} {\bibfnamefont {M.}~\bibnamefont
  {Adolph}}, \bibinfo {author} {\bibfnamefont {A.}~\bibnamefont {Aquila}},
  \bibinfo {author} {\bibfnamefont {S.}~\bibnamefont {Bajt}}, \bibinfo {author}
  {\bibfnamefont {A.}~\bibnamefont {Barty}}, \bibinfo {author} {\bibfnamefont
  {C.}~\bibnamefont {Bostedt}}, \bibinfo {author} {\bibfnamefont
  {J.}~\bibnamefont {Bozek}}, \bibinfo {author} {\bibfnamefont
  {C.}~\bibnamefont {Caleman}}, \bibinfo {author} {\bibfnamefont
  {R.}~\bibnamefont {Coffee}}, \bibinfo {author} {\bibfnamefont
  {N.}~\bibnamefont {Coppola}}, \bibinfo {author} {\bibfnamefont
  {T.}~\bibnamefont {Delmas}}, \bibinfo {author} {\bibfnamefont
  {S.}~\bibnamefont {Epp}}, \bibinfo {author} {\bibfnamefont {B.}~\bibnamefont
  {Erk}}, \bibinfo {author} {\bibfnamefont {L.}~\bibnamefont {Foucar}},
  \bibinfo {author} {\bibfnamefont {T.}~\bibnamefont {Gorkhover}}, \bibinfo
  {author} {\bibfnamefont {L.}~\bibnamefont {Gumprecht}}, \bibinfo {author}
  {\bibfnamefont {A.}~\bibnamefont {Hartmann}}, \bibinfo {author}
  {\bibfnamefont {R.}~\bibnamefont {Hartmann}}, \bibinfo {author}
  {\bibfnamefont {G.}~\bibnamefont {Hauser}}, \bibinfo {author} {\bibfnamefont
  {P.}~\bibnamefont {Holl}}, \bibinfo {author} {\bibfnamefont {A.}~\bibnamefont
  {H{\"o}mke}}, \bibinfo {author} {\bibfnamefont {N.}~\bibnamefont {Kimmel}},
  \bibinfo {author} {\bibfnamefont {F.}~\bibnamefont {Krasniqi}}, \bibinfo
  {author} {\bibfnamefont {K.-U.}\ \bibnamefont {K{\"u}hnel}}, \bibinfo
  {author} {\bibfnamefont {J.}~\bibnamefont {Maurer}}, \bibinfo {author}
  {\bibfnamefont {M.}~\bibnamefont {Messerschmidt}}, \bibinfo {author}
  {\bibfnamefont {R.}~\bibnamefont {Moshammer}}, \bibinfo {author}
  {\bibfnamefont {C.}~\bibnamefont {Reich}}, \bibinfo {author} {\bibfnamefont
  {B.}~\bibnamefont {Rudek}}, \bibinfo {author} {\bibfnamefont
  {R.}~\bibnamefont {Santra}}, \bibinfo {author} {\bibfnamefont
  {I.}~\bibnamefont {Schlichting}}, \bibinfo {author} {\bibfnamefont
  {C.}~\bibnamefont {Schmidt}}, \bibinfo {author} {\bibfnamefont
  {S.}~\bibnamefont {Schorb}}, \bibinfo {author} {\bibfnamefont
  {J.}~\bibnamefont {Schulz}}, \bibinfo {author} {\bibfnamefont
  {H.}~\bibnamefont {Soltau}}, \bibinfo {author} {\bibfnamefont {J.~C.~H.}\
  \bibnamefont {Spence}}, \bibinfo {author} {\bibfnamefont {D.}~\bibnamefont
  {Starodub}}, \bibinfo {author} {\bibfnamefont {L.}~\bibnamefont
  {Str{\"u}der}}, \bibinfo {author} {\bibfnamefont {J.}~\bibnamefont
  {Th{\o}gersen}}, \bibinfo {author} {\bibfnamefont {M.~J.~J.}\ \bibnamefont
  {Vrakking}}, \bibinfo {author} {\bibfnamefont {G.}~\bibnamefont
  {Weidenspointner}}, \bibinfo {author} {\bibfnamefont {T.~A.}\ \bibnamefont
  {White}}, \bibinfo {author} {\bibfnamefont {C.}~\bibnamefont {Wunderer}},
  \bibinfo {author} {\bibfnamefont {G.}~\bibnamefont {Meijer}}, \bibinfo
  {author} {\bibfnamefont {J.}~\bibnamefont {Ullrich}}, \bibinfo {author}
  {\bibfnamefont {H.}~\bibnamefont {Stapelfeldt}}, \bibinfo {author}
  {\bibfnamefont {D.}~\bibnamefont {Rolles}}, \ and\ \bibinfo {author}
  {\bibfnamefont {H.~N.}\ \bibnamefont {Chapman}},\ }\bibfield  {title}
  {\enquote {\bibinfo {title} {X-ray diffraction from isolated and strongly
  aligned gas-phase molecules with a free-electron laser},}\ }\href {\doibase
  10.1103/PhysRevLett.112.083002} {\bibfield  {journal} {\bibinfo  {journal}
  {Phys.\ Rev.\ Lett.}\ }\textbf {\bibinfo {volume} {112}},\ \bibinfo {pages}
  {083002} (\bibinfo {year} {2014})},\ \Eprint {http://arxiv.org/abs/1307.4577}
  {arXiv:1307.4577 [physics]} \BibitemShut {NoStop}%
\bibitem [{\citenamefont {Glownia}\ \emph {et~al.}(2016)\citenamefont
  {Glownia}, \citenamefont {Natan}, \citenamefont {Cryan}, \citenamefont
  {Hartsock}, \citenamefont {Kozina}, \citenamefont {Minitti}, \citenamefont
  {Nelson}, \citenamefont {Robinson}, \citenamefont {Sato}, \citenamefont {van
  Driel}, \citenamefont {Welch}, \citenamefont {Weninger}, \citenamefont
  {Zhu},\ and\ \citenamefont {Bucksbaum}}]{Glownia:PRL117:153003}%
  \BibitemOpen
  \bibfield  {author} {\bibinfo {author} {\bibfnamefont {J.~M.}\ \bibnamefont
  {Glownia}}, \bibinfo {author} {\bibfnamefont {A.}~\bibnamefont {Natan}},
  \bibinfo {author} {\bibfnamefont {J.~P.}\ \bibnamefont {Cryan}}, \bibinfo
  {author} {\bibfnamefont {R.}~\bibnamefont {Hartsock}}, \bibinfo {author}
  {\bibfnamefont {M.}~\bibnamefont {Kozina}}, \bibinfo {author} {\bibfnamefont
  {M.~P.}\ \bibnamefont {Minitti}}, \bibinfo {author} {\bibfnamefont
  {S.}~\bibnamefont {Nelson}}, \bibinfo {author} {\bibfnamefont
  {J.}~\bibnamefont {Robinson}}, \bibinfo {author} {\bibfnamefont
  {T.}~\bibnamefont {Sato}}, \bibinfo {author} {\bibfnamefont {T.}~\bibnamefont
  {van Driel}}, \bibinfo {author} {\bibfnamefont {G.}~\bibnamefont {Welch}},
  \bibinfo {author} {\bibfnamefont {C.}~\bibnamefont {Weninger}}, \bibinfo
  {author} {\bibfnamefont {D.}~\bibnamefont {Zhu}}, \ and\ \bibinfo {author}
  {\bibfnamefont {P.~H.}\ \bibnamefont {Bucksbaum}},\ }\bibfield  {title}
  {\enquote {\bibinfo {title} {Self-referenced coherent diffraction x-ray movie
  of \aa{}ngstrom- and femtosecond-scale atomic motion},}\ }\href {\doibase
  10.1103/PhysRevLett.117.153003} {\bibfield  {journal} {\bibinfo  {journal}
  {Phys.\ Rev.\ Lett.}\ }\textbf {\bibinfo {volume} {117}},\ \bibinfo {pages}
  {153003} (\bibinfo {year} {2016})},\ \Eprint
  {http://arxiv.org/abs/1608.03039} {arXiv:1608.03039 [physics]} \BibitemShut
  {NoStop}%
\bibitem [{\citenamefont {Weathersby}\ \emph {et~al.}(2015)\citenamefont
  {Weathersby}, \citenamefont {Brown}, \citenamefont {Centurion}, \citenamefont
  {Chase}, \citenamefont {Coffee}, \citenamefont {Corbett}, \citenamefont
  {Eichner}, \citenamefont {Frisch}, \citenamefont {Fry}, \citenamefont
  {G{\"u}hr}, \citenamefont {Hartmann}, \citenamefont {Hast}, \citenamefont
  {Hettel}, \citenamefont {Jobe}, \citenamefont {Jongewaard}, \citenamefont
  {Lewandowski}, \citenamefont {Li}, \citenamefont {Lindenberg}, \citenamefont
  {Makasyuk}, \citenamefont {May}, \citenamefont {McCormick}, \citenamefont
  {Nguyen}, \citenamefont {Reid}, \citenamefont {Shen}, \citenamefont
  {Sokolowski-Tinten}, \citenamefont {Vecchione}, \citenamefont {Vetter},
  \citenamefont {Wu}, \citenamefont {Yang}, \citenamefont {D{\"u}rr},\ and\
  \citenamefont {Wang}}]{Weathersby:RSI86:073702}%
  \BibitemOpen
  \bibfield  {author} {\bibinfo {author} {\bibfnamefont {S.~P.}\ \bibnamefont
  {Weathersby}}, \bibinfo {author} {\bibfnamefont {G.}~\bibnamefont {Brown}},
  \bibinfo {author} {\bibfnamefont {M.}~\bibnamefont {Centurion}}, \bibinfo
  {author} {\bibfnamefont {T.~F.}\ \bibnamefont {Chase}}, \bibinfo {author}
  {\bibfnamefont {R.}~\bibnamefont {Coffee}}, \bibinfo {author} {\bibfnamefont
  {J.}~\bibnamefont {Corbett}}, \bibinfo {author} {\bibfnamefont {J.~P.}\
  \bibnamefont {Eichner}}, \bibinfo {author} {\bibfnamefont {J.~C.}\
  \bibnamefont {Frisch}}, \bibinfo {author} {\bibfnamefont {A.~R.}\
  \bibnamefont {Fry}}, \bibinfo {author} {\bibfnamefont {M.}~\bibnamefont
  {G{\"u}hr}}, \bibinfo {author} {\bibfnamefont {N.}~\bibnamefont {Hartmann}},
  \bibinfo {author} {\bibfnamefont {C.}~\bibnamefont {Hast}}, \bibinfo {author}
  {\bibfnamefont {R.}~\bibnamefont {Hettel}}, \bibinfo {author} {\bibfnamefont
  {R.~K.}\ \bibnamefont {Jobe}}, \bibinfo {author} {\bibfnamefont {E.~N.}\
  \bibnamefont {Jongewaard}}, \bibinfo {author} {\bibfnamefont {J.~R.}\
  \bibnamefont {Lewandowski}}, \bibinfo {author} {\bibfnamefont {R.~K.}\
  \bibnamefont {Li}}, \bibinfo {author} {\bibfnamefont {A.~M.}\ \bibnamefont
  {Lindenberg}}, \bibinfo {author} {\bibfnamefont {I.}~\bibnamefont
  {Makasyuk}}, \bibinfo {author} {\bibfnamefont {J.~E.}\ \bibnamefont {May}},
  \bibinfo {author} {\bibfnamefont {D.}~\bibnamefont {McCormick}}, \bibinfo
  {author} {\bibfnamefont {M.~N.}\ \bibnamefont {Nguyen}}, \bibinfo {author}
  {\bibfnamefont {A.~H.}\ \bibnamefont {Reid}}, \bibinfo {author}
  {\bibfnamefont {X.}~\bibnamefont {Shen}}, \bibinfo {author} {\bibfnamefont
  {K.}~\bibnamefont {Sokolowski-Tinten}}, \bibinfo {author} {\bibfnamefont
  {T.}~\bibnamefont {Vecchione}}, \bibinfo {author} {\bibfnamefont {S.~L.}\
  \bibnamefont {Vetter}}, \bibinfo {author} {\bibfnamefont {J.}~\bibnamefont
  {Wu}}, \bibinfo {author} {\bibfnamefont {J.}~\bibnamefont {Yang}}, \bibinfo
  {author} {\bibfnamefont {H.~A.}\ \bibnamefont {D{\"u}rr}}, \ and\ \bibinfo
  {author} {\bibfnamefont {X.~J.}\ \bibnamefont {Wang}},\ }\bibfield  {title}
  {\enquote {\bibinfo {title} {Mega-electron-volt ultrafast electron
  diffraction at {SLAC} {N}ational {A}ccelerator {L}aboratory},}\ }\href
  {\doibase 10.1063/1.4926994} {\bibfield  {journal} {\bibinfo  {journal}
  {Rev.\ Sci.\ Instrum.}\ }\textbf {\bibinfo {volume} {86}},\ \bibinfo {pages}
  {073702} (\bibinfo {year} {2015})}\BibitemShut {NoStop}%
\bibitem [{\citenamefont {Yang}\ \emph {et~al.}(2016)\citenamefont {Yang},
  \citenamefont {Guehr}, \citenamefont {Shen}, \citenamefont {Li},
  \citenamefont {Vecchione}, \citenamefont {Coffee}, \citenamefont {Corbett},
  \citenamefont {Fry}, \citenamefont {Hartmann}, \citenamefont {Hast},
  \citenamefont {Hegazy}, \citenamefont {Jobe}, \citenamefont {Makasyuk},
  \citenamefont {Robinson}, \citenamefont {Robinson}, \citenamefont {Vetter},
  \citenamefont {Weathersby}, \citenamefont {Yoneda}, \citenamefont {Wang},\
  and\ \citenamefont {Centurion}}]{Yang:PRL117:153002}%
  \BibitemOpen
  \bibfield  {author} {\bibinfo {author} {\bibfnamefont {J.}~\bibnamefont
  {Yang}}, \bibinfo {author} {\bibfnamefont {M.}~\bibnamefont {Guehr}},
  \bibinfo {author} {\bibfnamefont {X.}~\bibnamefont {Shen}}, \bibinfo {author}
  {\bibfnamefont {R.}~\bibnamefont {Li}}, \bibinfo {author} {\bibfnamefont
  {T.}~\bibnamefont {Vecchione}}, \bibinfo {author} {\bibfnamefont
  {R.}~\bibnamefont {Coffee}}, \bibinfo {author} {\bibfnamefont
  {J.}~\bibnamefont {Corbett}}, \bibinfo {author} {\bibfnamefont
  {A.}~\bibnamefont {Fry}}, \bibinfo {author} {\bibfnamefont {N.}~\bibnamefont
  {Hartmann}}, \bibinfo {author} {\bibfnamefont {C.}~\bibnamefont {Hast}},
  \bibinfo {author} {\bibfnamefont {K.}~\bibnamefont {Hegazy}}, \bibinfo
  {author} {\bibfnamefont {K.}~\bibnamefont {Jobe}}, \bibinfo {author}
  {\bibfnamefont {I.}~\bibnamefont {Makasyuk}}, \bibinfo {author}
  {\bibfnamefont {J.}~\bibnamefont {Robinson}}, \bibinfo {author}
  {\bibfnamefont {M.~S.}\ \bibnamefont {Robinson}}, \bibinfo {author}
  {\bibfnamefont {S.}~\bibnamefont {Vetter}}, \bibinfo {author} {\bibfnamefont
  {S.}~\bibnamefont {Weathersby}}, \bibinfo {author} {\bibfnamefont
  {C.}~\bibnamefont {Yoneda}}, \bibinfo {author} {\bibfnamefont
  {X.}~\bibnamefont {Wang}}, \ and\ \bibinfo {author} {\bibfnamefont
  {M.}~\bibnamefont {Centurion}},\ }\bibfield  {title} {\enquote {\bibinfo
  {title} {Diffractive imaging of coherent nuclear motion in isolated
  molecules},}\ }\href {\doibase 10.1103/PhysRevLett.117.153002} {\bibfield
  {journal} {\bibinfo  {journal} {Phys.\ Rev.\ Lett.}\ }\textbf {\bibinfo
  {volume} {117}},\ \bibinfo {pages} {153002} (\bibinfo {year}
  {2016})}\BibitemShut {NoStop}%
\bibitem [{\citenamefont {Corkum}(1993)}]{Corkum:PRL71:1994}%
  \BibitemOpen
  \bibfield  {author} {\bibinfo {author} {\bibfnamefont {P.~B.}\ \bibnamefont
  {Corkum}},\ }\bibfield  {title} {\enquote {\bibinfo {title} {Plasma
  perspective on strong-field multiphoton ionization},}\ }\href {\doibase
  10.1103/PhysRevLett.71.1994} {\bibfield  {journal} {\bibinfo  {journal}
  {Phys.\ Rev.\ Lett.}\ }\textbf {\bibinfo {volume} {71}},\ \bibinfo {pages}
  {1994--1997} (\bibinfo {year} {1993})}\BibitemShut {NoStop}%
\bibitem [{\citenamefont {Spanner}\ \emph {et~al.}(2004)\citenamefont
  {Spanner}, \citenamefont {Smirnova}, \citenamefont {Corkum},\ and\
  \citenamefont {Ivanov}}]{Spanner:JPB37:L243}%
  \BibitemOpen
  \bibfield  {author} {\bibinfo {author} {\bibfnamefont {M.}~\bibnamefont
  {Spanner}}, \bibinfo {author} {\bibfnamefont {O.}~\bibnamefont {Smirnova}},
  \bibinfo {author} {\bibfnamefont {P.~B.}\ \bibnamefont {Corkum}}, \ and\
  \bibinfo {author} {\bibfnamefont {M.~Y.}\ \bibnamefont {Ivanov}},\ }\bibfield
   {title} {\enquote {\bibinfo {title} {Reading diffraction images in strong
  field ionization of diatomic molecules},}\ }\href {\doibase
  10.1088/0953-4075/37/12/L02} {\bibfield  {journal} {\bibinfo  {journal} {J.\
  Phys.\ B}\ }\textbf {\bibinfo {volume} {37}},\ \bibinfo {pages} {L243--L250}
  (\bibinfo {year} {2004})}\BibitemShut {NoStop}%
\bibitem [{\citenamefont {Blaga}\ \emph {et~al.}(2012)\citenamefont {Blaga},
  \citenamefont {Xu}, \citenamefont {DiChiara}, \citenamefont {Sistrunk},
  \citenamefont {Zhang}, \citenamefont {Agostini}, \citenamefont {Miller},
  \citenamefont {DiMauro},\ and\ \citenamefont {Lin}}]{Blaga:Nature483:194}%
  \BibitemOpen
  \bibfield  {author} {\bibinfo {author} {\bibfnamefont {C.~I.}\ \bibnamefont
  {Blaga}}, \bibinfo {author} {\bibfnamefont {J.}~\bibnamefont {Xu}}, \bibinfo
  {author} {\bibfnamefont {A.~D.}\ \bibnamefont {DiChiara}}, \bibinfo {author}
  {\bibfnamefont {E.}~\bibnamefont {Sistrunk}}, \bibinfo {author}
  {\bibfnamefont {K.}~\bibnamefont {Zhang}}, \bibinfo {author} {\bibfnamefont
  {P.}~\bibnamefont {Agostini}}, \bibinfo {author} {\bibfnamefont {T.~A.}\
  \bibnamefont {Miller}}, \bibinfo {author} {\bibfnamefont {L.~F.}\
  \bibnamefont {DiMauro}}, \ and\ \bibinfo {author} {\bibfnamefont {C.~D.}\
  \bibnamefont {Lin}},\ }\bibfield  {title} {\enquote {\bibinfo {title}
  {Imaging ultrafast molecular dynamics with laser-induced electron
  diffraction},}\ }\href {\doibase 10.1038/nature10820} {\bibfield  {journal}
  {\bibinfo  {journal} {Nature}\ }\textbf {\bibinfo {volume} {483}},\ \bibinfo
  {pages} {194--197} (\bibinfo {year} {2012})}\BibitemShut {NoStop}%
\bibitem [{\citenamefont {Okunishi}\ \emph {et~al.}(2008)\citenamefont
  {Okunishi}, \citenamefont {Morishita}, \citenamefont {Pr\"umper},
  \citenamefont {Shimada}, \citenamefont {Lin}, \citenamefont {Watanabe},\ and\
  \citenamefont {Ueda}}]{Okunishi:PRL100:143001}%
  \BibitemOpen
  \bibfield  {author} {\bibinfo {author} {\bibfnamefont {M.}~\bibnamefont
  {Okunishi}}, \bibinfo {author} {\bibfnamefont {T.}~\bibnamefont {Morishita}},
  \bibinfo {author} {\bibfnamefont {G.}~\bibnamefont {Pr\"umper}}, \bibinfo
  {author} {\bibfnamefont {K.}~\bibnamefont {Shimada}}, \bibinfo {author}
  {\bibfnamefont {C.~D.}\ \bibnamefont {Lin}}, \bibinfo {author} {\bibfnamefont
  {S.}~\bibnamefont {Watanabe}}, \ and\ \bibinfo {author} {\bibfnamefont
  {K.}~\bibnamefont {Ueda}},\ }\bibfield  {title} {\enquote {\bibinfo {title}
  {Experimental retrieval of target structure information from laser-induced
  rescattered photoelectron momentum distributions},}\ }\href {\doibase
  10.1103/PhysRevLett.100.143001} {\bibfield  {journal} {\bibinfo  {journal}
  {Phys.\ Rev.\ Lett.}\ }\textbf {\bibinfo {volume} {100}},\ \bibinfo {pages}
  {143001} (\bibinfo {year} {2008})}\BibitemShut {NoStop}%
\bibitem [{\citenamefont {Xu}\ \emph {et~al.}(2012)\citenamefont {Xu},
  \citenamefont {Blaga}, \citenamefont {DiChiara}, \citenamefont {Sistrunk},
  \citenamefont {Zhang}, \citenamefont {Chen}, \citenamefont {Le},
  \citenamefont {Morishita}, \citenamefont {Lin}, \citenamefont {Agostini},\
  and\ \citenamefont {DiMauro}}]{Xu:PRL109:233002}%
  \BibitemOpen
  \bibfield  {author} {\bibinfo {author} {\bibfnamefont {J.}~\bibnamefont
  {Xu}}, \bibinfo {author} {\bibfnamefont {C.~I.}\ \bibnamefont {Blaga}},
  \bibinfo {author} {\bibfnamefont {A.~D.}\ \bibnamefont {DiChiara}}, \bibinfo
  {author} {\bibfnamefont {E.}~\bibnamefont {Sistrunk}}, \bibinfo {author}
  {\bibfnamefont {K.}~\bibnamefont {Zhang}}, \bibinfo {author} {\bibfnamefont
  {Z.}~\bibnamefont {Chen}}, \bibinfo {author} {\bibfnamefont {A.-T.}\
  \bibnamefont {Le}}, \bibinfo {author} {\bibfnamefont {T.}~\bibnamefont
  {Morishita}}, \bibinfo {author} {\bibfnamefont {C.~D.}\ \bibnamefont {Lin}},
  \bibinfo {author} {\bibfnamefont {P.}~\bibnamefont {Agostini}}, \ and\
  \bibinfo {author} {\bibfnamefont {L.~F.}\ \bibnamefont {DiMauro}},\
  }\bibfield  {title} {\enquote {\bibinfo {title} {Laser-induced electron
  diffraction for probing rare gas atoms},}\ }\href {\doibase
  10.1103/PhysRevLett.109.233002} {\bibfield  {journal} {\bibinfo  {journal}
  {Phys.\ Rev.\ Lett.}\ }\textbf {\bibinfo {volume} {109}},\ \bibinfo {pages}
  {233002} (\bibinfo {year} {2012})}\BibitemShut {NoStop}%
\bibitem [{\citenamefont {Xu}\ \emph {et~al.}(2014)\citenamefont {Xu},
  \citenamefont {Blaga}, \citenamefont {Zhang}, \citenamefont {Lai},
  \citenamefont {Lin}, \citenamefont {Miller}, \citenamefont {Agostini},\ and\
  \citenamefont {DiMauro}}]{Xu:NatComm5:4635}%
  \BibitemOpen
  \bibfield  {author} {\bibinfo {author} {\bibfnamefont {J.}~\bibnamefont
  {Xu}}, \bibinfo {author} {\bibfnamefont {C.~I.}\ \bibnamefont {Blaga}},
  \bibinfo {author} {\bibfnamefont {K.}~\bibnamefont {Zhang}}, \bibinfo
  {author} {\bibfnamefont {Y.~H.}\ \bibnamefont {Lai}}, \bibinfo {author}
  {\bibfnamefont {C.~D.}\ \bibnamefont {Lin}}, \bibinfo {author} {\bibfnamefont
  {T.~A.}\ \bibnamefont {Miller}}, \bibinfo {author} {\bibfnamefont
  {P.}~\bibnamefont {Agostini}}, \ and\ \bibinfo {author} {\bibfnamefont
  {L.~F.}\ \bibnamefont {DiMauro}},\ }\bibfield  {title} {\enquote {\bibinfo
  {title} {Diffraction using laser-driven broadband electron wave packets},}\
  }\href {\doibase 10.1038/ncomms5635} {\bibfield  {journal} {\bibinfo
  {journal} {Nat. Commun.}\ }\textbf {\bibinfo {volume} {5}},\ \bibinfo {pages}
  {4635} (\bibinfo {year} {2014})}\BibitemShut {NoStop}%
\bibitem [{\citenamefont {Pullen}\ \emph {et~al.}(2015)\citenamefont {Pullen},
  \citenamefont {Wolter}, \citenamefont {Le}, \citenamefont {Baudisch},
  \citenamefont {Hemmer}, \citenamefont {Senftleben}, \citenamefont {Schroter},
  \citenamefont {Ullrich}, \citenamefont {Moshammer}, \citenamefont {Lin},\
  and\ \citenamefont {Biegert}}]{Pullen:NatComm6:7262}%
  \BibitemOpen
  \bibfield  {author} {\bibinfo {author} {\bibfnamefont {M.~G.}\ \bibnamefont
  {Pullen}}, \bibinfo {author} {\bibfnamefont {B.}~\bibnamefont {Wolter}},
  \bibinfo {author} {\bibfnamefont {A.-T.}\ \bibnamefont {Le}}, \bibinfo
  {author} {\bibfnamefont {M.}~\bibnamefont {Baudisch}}, \bibinfo {author}
  {\bibfnamefont {M.}~\bibnamefont {Hemmer}}, \bibinfo {author} {\bibfnamefont
  {A.}~\bibnamefont {Senftleben}}, \bibinfo {author} {\bibfnamefont {C.~D.}\
  \bibnamefont {Schroter}}, \bibinfo {author} {\bibfnamefont {J.}~\bibnamefont
  {Ullrich}}, \bibinfo {author} {\bibfnamefont {R.}~\bibnamefont {Moshammer}},
  \bibinfo {author} {\bibfnamefont {C.~D.}\ \bibnamefont {Lin}}, \ and\
  \bibinfo {author} {\bibfnamefont {J.}~\bibnamefont {Biegert}},\ }\bibfield
  {title} {\enquote {\bibinfo {title} {Imaging an aligned polyatomic molecule
  with laser-induced electron diffraction},}\ }\href {\doibase
  10.1038/ncomms8262} {\bibfield  {journal} {\bibinfo  {journal} {Nat.
  Commun.}\ }\textbf {\bibinfo {volume} {6}},\ \bibinfo {pages} {7262}
  (\bibinfo {year} {2015})}\BibitemShut {NoStop}%
\bibitem [{\citenamefont {Ito}\ \emph {et~al.}(2017)\citenamefont {Ito},
  \citenamefont {Carranza}, \citenamefont {Okunishi}, \citenamefont
  {Lucchese},\ and\ \citenamefont {Ueda}}]{Ito:PRA96:053414}%
  \BibitemOpen
  \bibfield  {author} {\bibinfo {author} {\bibfnamefont {Y.}~\bibnamefont
  {Ito}}, \bibinfo {author} {\bibfnamefont {R.}~\bibnamefont {Carranza}},
  \bibinfo {author} {\bibfnamefont {M.}~\bibnamefont {Okunishi}}, \bibinfo
  {author} {\bibfnamefont {R.~R.}\ \bibnamefont {Lucchese}}, \ and\ \bibinfo
  {author} {\bibfnamefont {K.}~\bibnamefont {Ueda}},\ }\bibfield  {title}
  {\enquote {\bibinfo {title} {Extraction of geometrical structure of ethylene
  molecules by laser-induced electron diffraction combined with ab initio
  scattering calculations},}\ }\href {\doibase 10.1103/PhysRevA.96.053414}
  {\bibfield  {journal} {\bibinfo  {journal} {Phys.\ Rev.\ A}\ }\textbf
  {\bibinfo {volume} {96}},\ \bibinfo {pages} {053414} (\bibinfo {year}
  {2017})}\BibitemShut {NoStop}%
\bibitem [{\citenamefont {Ito}\ \emph {et~al.}(2016)\citenamefont {Ito},
  \citenamefont {Wang}, \citenamefont {Le}, \citenamefont {Okunishi},
  \citenamefont {Ding}, \citenamefont {Lin},\ and\ \citenamefont
  {Ueda}}]{Ito:SD3:034303}%
  \BibitemOpen
  \bibfield  {author} {\bibinfo {author} {\bibfnamefont {Y.}~\bibnamefont
  {Ito}}, \bibinfo {author} {\bibfnamefont {C.}~\bibnamefont {Wang}}, \bibinfo
  {author} {\bibfnamefont {A.-T.}\ \bibnamefont {Le}}, \bibinfo {author}
  {\bibfnamefont {M.}~\bibnamefont {Okunishi}}, \bibinfo {author}
  {\bibfnamefont {D.}~\bibnamefont {Ding}}, \bibinfo {author} {\bibfnamefont
  {C.~D.}\ \bibnamefont {Lin}}, \ and\ \bibinfo {author} {\bibfnamefont
  {K.}~\bibnamefont {Ueda}},\ }\bibfield  {title} {\enquote {\bibinfo {title}
  {Extracting conformational structure information of benzene molecules via
  laser-induced electron diffraction},}\ }\href {\doibase 10.1063/1.4952602}
  {\bibfield  {journal} {\bibinfo  {journal} {Struct.\ Dyn.}\ }\textbf
  {\bibinfo {volume} {3}},\ \bibinfo {pages} {034303} (\bibinfo {year}
  {2016})}\BibitemShut {NoStop}%
\bibitem [{\citenamefont {Walt}\ \emph {et~al.}(2017)\citenamefont {Walt},
  \citenamefont {Ram}, \citenamefont {Atala}, \citenamefont
  {Shvetsov-Shilovski}, \citenamefont {von Conta}, \citenamefont {Baykusheva},
  \citenamefont {Lein},\ and\ \citenamefont
  {W\"{o}rner}}]{Walt:NatComm8:15651}%
  \BibitemOpen
  \bibfield  {author} {\bibinfo {author} {\bibfnamefont {S.~G.}\ \bibnamefont
  {Walt}}, \bibinfo {author} {\bibfnamefont {B.~N.}\ \bibnamefont {Ram}},
  \bibinfo {author} {\bibfnamefont {M.}~\bibnamefont {Atala}}, \bibinfo
  {author} {\bibfnamefont {N.~I.}\ \bibnamefont {Shvetsov-Shilovski}}, \bibinfo
  {author} {\bibfnamefont {A.}~\bibnamefont {von Conta}}, \bibinfo {author}
  {\bibfnamefont {D.}~\bibnamefont {Baykusheva}}, \bibinfo {author}
  {\bibfnamefont {M.}~\bibnamefont {Lein}}, \ and\ \bibinfo {author}
  {\bibfnamefont {H.~J.}\ \bibnamefont {W\"{o}rner}},\ }\bibfield  {title}
  {\enquote {\bibinfo {title} {Dynamics of valence-shell electrons and nuclei
  probed by strong-field holography and rescattering},}\ }\href {\doibase
  10.1038/ncomms15651} {\bibfield  {journal} {\bibinfo  {journal} {Nat.
  Commun.}\ }\textbf {\bibinfo {volume} {8}},\ \bibinfo {pages} {15651}
  (\bibinfo {year} {2017})}\BibitemShut {NoStop}%
\bibitem [{\citenamefont {Wolter}\ \emph {et~al.}(2016)\citenamefont {Wolter},
  \citenamefont {Pullen}, \citenamefont {Le}, \citenamefont {Baudisch},
  \citenamefont {Doblhoff-Dier}, \citenamefont {Senftleben}, \citenamefont
  {Hemmer}, \citenamefont {Schroter}, \citenamefont {Ullrich}, \citenamefont
  {Pfeifer}, \citenamefont {Moshammer}, \citenamefont {Gr{\"a}fe},
  \citenamefont {Vendrell}, \citenamefont {Lin},\ and\ \citenamefont
  {Biegert}}]{Wolter:Science354:308}%
  \BibitemOpen
  \bibfield  {author} {\bibinfo {author} {\bibfnamefont {B.}~\bibnamefont
  {Wolter}}, \bibinfo {author} {\bibfnamefont {M.~G.}\ \bibnamefont {Pullen}},
  \bibinfo {author} {\bibfnamefont {A.~T.}\ \bibnamefont {Le}}, \bibinfo
  {author} {\bibfnamefont {M.}~\bibnamefont {Baudisch}}, \bibinfo {author}
  {\bibfnamefont {K.}~\bibnamefont {Doblhoff-Dier}}, \bibinfo {author}
  {\bibfnamefont {A.}~\bibnamefont {Senftleben}}, \bibinfo {author}
  {\bibfnamefont {M.}~\bibnamefont {Hemmer}}, \bibinfo {author} {\bibfnamefont
  {C.~D.}\ \bibnamefont {Schroter}}, \bibinfo {author} {\bibfnamefont
  {J.}~\bibnamefont {Ullrich}}, \bibinfo {author} {\bibfnamefont
  {T.}~\bibnamefont {Pfeifer}}, \bibinfo {author} {\bibfnamefont
  {R.}~\bibnamefont {Moshammer}}, \bibinfo {author} {\bibfnamefont
  {S.}~\bibnamefont {Gr{\"a}fe}}, \bibinfo {author} {\bibfnamefont
  {O.}~\bibnamefont {Vendrell}}, \bibinfo {author} {\bibfnamefont {C.~D.}\
  \bibnamefont {Lin}}, \ and\ \bibinfo {author} {\bibfnamefont
  {J.}~\bibnamefont {Biegert}},\ }\bibfield  {title} {\enquote {\bibinfo
  {title} {Ultrafast electron diffraction imaging of bond breaking in
  di-ionized acetylene},}\ }\href {\doibase 10.1126/science.aah3429} {\bibfield
   {journal} {\bibinfo  {journal} {Science}\ }\textbf {\bibinfo {volume}
  {354}},\ \bibinfo {pages} {308--312} (\bibinfo {year} {2016})}\BibitemShut
  {NoStop}%
\bibitem [{\citenamefont {Brouard}\ \emph {et~al.}(2007)\citenamefont
  {Brouard}, \citenamefont {Green}, \citenamefont {Quadrini},\ and\
  \citenamefont {Vallance}}]{Brouard:JCP127:084304}%
  \BibitemOpen
  \bibfield  {author} {\bibinfo {author} {\bibfnamefont {M.}~\bibnamefont
  {Brouard}}, \bibinfo {author} {\bibfnamefont {A.~V.}\ \bibnamefont {Green}},
  \bibinfo {author} {\bibfnamefont {F.}~\bibnamefont {Quadrini}}, \ and\
  \bibinfo {author} {\bibfnamefont {C.}~\bibnamefont {Vallance}},\ }\bibfield
  {title} {\enquote {\bibinfo {title} {Photodissociation dynamics of {OCS} at
  248~nm: The s(d21) atomic angular momentum polarization},}\ }\href {\doibase
  10.1063/1.2757618} {\bibfield  {journal} {\bibinfo  {journal} {J.\ Chem.\
  Phys.}\ }\textbf {\bibinfo {volume} {127}},\ \bibinfo {pages} {084304}
  (\bibinfo {year} {2007})}\BibitemShut {NoStop}%
\bibitem [{\citenamefont {Suzuki}\ \emph {et~al.}(1998)\citenamefont {Suzuki},
  \citenamefont {Katayanagi}, \citenamefont {Nanbu},\ and\ \citenamefont
  {Aoyagi}}]{Suzuki:JCP109:5778}%
  \BibitemOpen
  \bibfield  {author} {\bibinfo {author} {\bibfnamefont {T.}~\bibnamefont
  {Suzuki}}, \bibinfo {author} {\bibfnamefont {H.}~\bibnamefont {Katayanagi}},
  \bibinfo {author} {\bibfnamefont {S.}~\bibnamefont {Nanbu}}, \ and\ \bibinfo
  {author} {\bibfnamefont {M.}~\bibnamefont {Aoyagi}},\ }\bibfield  {title}
  {\enquote {\bibinfo {title} {Nonadiabatic bending dissociation in 16 valence
  electron system {OCS}},}\ }\href {\doibase 10.1063/1.477200} {\bibfield
  {journal} {\bibinfo  {journal} {J.\ Chem.\ Phys.}\ }\textbf {\bibinfo
  {volume} {109}},\ \bibinfo {pages} {5778--5794} (\bibinfo {year}
  {1998})}\BibitemShut {NoStop}%
\bibitem [{\citenamefont {Sivakumar}\ \emph {et~al.}(1985)\citenamefont
  {Sivakumar}, \citenamefont {Burak}, \citenamefont {Cheung}, \citenamefont
  {Houston},\ and\ \citenamefont {Hepburn}}]{Sivakumar:JPC89:3609}%
  \BibitemOpen
  \bibfield  {author} {\bibinfo {author} {\bibfnamefont {N.}~\bibnamefont
  {Sivakumar}}, \bibinfo {author} {\bibfnamefont {I.}~\bibnamefont {Burak}},
  \bibinfo {author} {\bibfnamefont {W.~Y.}\ \bibnamefont {Cheung}}, \bibinfo
  {author} {\bibfnamefont {P.~L.}\ \bibnamefont {Houston}}, \ and\ \bibinfo
  {author} {\bibfnamefont {J.~W.}\ \bibnamefont {Hepburn}},\ }\bibfield
  {title} {\enquote {\bibinfo {title} {State-resolved photofragmentation of
  carbonyl sulfide ({OCS}) monomers and clusters},}\ }\href {\doibase
  10.1021/j100263a008} {\bibfield  {journal} {\bibinfo  {journal} {J.\ Phys.\
  Chem.}\ }\textbf {\bibinfo {volume} {89}},\ \bibinfo {pages} {3609--3611}
  (\bibinfo {year} {1985})}\BibitemShut {NoStop}%
\bibitem [{\citenamefont {Sanderson}\ \emph {et~al.}(2002)\citenamefont
  {Sanderson}, \citenamefont {Goodworth}, \citenamefont {El-Zein},
  \citenamefont {Bryan}, \citenamefont {Newell}, \citenamefont {Langley},\ and\
  \citenamefont {Taday}}]{Sanderson:PRA65:043403}%
  \BibitemOpen
  \bibfield  {author} {\bibinfo {author} {\bibfnamefont {J.~H.}\ \bibnamefont
  {Sanderson}}, \bibinfo {author} {\bibfnamefont {T.~R.~J.}\ \bibnamefont
  {Goodworth}}, \bibinfo {author} {\bibfnamefont {A.}~\bibnamefont {El-Zein}},
  \bibinfo {author} {\bibfnamefont {W.~A.}\ \bibnamefont {Bryan}}, \bibinfo
  {author} {\bibfnamefont {W.~R.}\ \bibnamefont {Newell}}, \bibinfo {author}
  {\bibfnamefont {A.~J.}\ \bibnamefont {Langley}}, \ and\ \bibinfo {author}
  {\bibfnamefont {P.~F.}\ \bibnamefont {Taday}},\ }\bibfield  {title} {\enquote
  {\bibinfo {title} {Coulombic and pre-coulombic geometry evolution of carbonyl
  sulfide in an intense femtosecond laser pulse, determined by momentum
  imaging},}\ }\href {\doibase 10.1103/PhysRevA.65.043403} {\bibfield
  {journal} {\bibinfo  {journal} {Phys.\ Rev.\ A}\ }\textbf {\bibinfo {volume}
  {65}},\ \bibinfo {pages} {043403} (\bibinfo {year} {2002})}\BibitemShut
  {NoStop}%
\bibitem [{\citenamefont {Eppink}\ and\ \citenamefont
  {Parker}(1997)}]{Eppink:RSI68:3477}%
  \BibitemOpen
  \bibfield  {author} {\bibinfo {author} {\bibfnamefont {A.~T. J.~B.}\
  \bibnamefont {Eppink}}\ and\ \bibinfo {author} {\bibfnamefont {D.~H.}\
  \bibnamefont {Parker}},\ }\bibfield  {title} {\enquote {\bibinfo {title}
  {Velocity map imaging of ions and electrons using electrostatic lenses:
  {A}pplication in photoelectron and photofragment ion imaging of molecular
  oxygen},}\ }\href {\doibase 10.1063/1.1148310} {\bibfield  {journal}
  {\bibinfo  {journal} {Rev.\ Sci.\ Instrum.}\ }\textbf {\bibinfo {volume}
  {68}},\ \bibinfo {pages} {3477--3484} (\bibinfo {year} {1997})}\BibitemShut
  {NoStop}%
\bibitem [{\citenamefont {Salvat}, \citenamefont {Jablonski},\ and\
  \citenamefont {Powell}(2005)}]{Salvat:CPC165:157}%
  \BibitemOpen
  \bibfield  {author} {\bibinfo {author} {\bibfnamefont {F.}~\bibnamefont
  {Salvat}}, \bibinfo {author} {\bibfnamefont {A.}~\bibnamefont {Jablonski}}, \
  and\ \bibinfo {author} {\bibfnamefont {C.}~\bibnamefont {Powell}},\
  }\bibfield  {title} {\enquote {\bibinfo {title} {{ELSEPA} -- {D}irac
  partial-wave calculation of elastic scattering of electrons and positrons by
  atoms, positive ions and molecules},}\ }\href {\doibase
  10.1016/j.cpc.2004.09.006} {\bibfield  {journal} {\bibinfo  {journal} {Comp.\
  Phys.\ Comm.}\ }\textbf {\bibinfo {volume} {165}},\ \bibinfo {pages}
  {157--190} (\bibinfo {year} {2005})}\BibitemShut {NoStop}%
\bibitem [{\citenamefont {Dakin}, \citenamefont {Good},\ and\ \citenamefont
  {Coles}(1947)}]{Dakin:PR71:640}%
  \BibitemOpen
  \bibfield  {author} {\bibinfo {author} {\bibfnamefont {T.~W.}\ \bibnamefont
  {Dakin}}, \bibinfo {author} {\bibfnamefont {W.~E.}\ \bibnamefont {Good}}, \
  and\ \bibinfo {author} {\bibfnamefont {D.~K.}\ \bibnamefont {Coles}},\
  }\bibfield  {title} {\enquote {\bibinfo {title} {Bond distances in {OCS} from
  microwave absorption lines},}\ }\href {\doibase 10.1103/PhysRev.71.640.2}
  {\bibfield  {journal} {\bibinfo  {journal} {Phys. Rev.}\ }\textbf {\bibinfo
  {volume} {71}},\ \bibinfo {pages} {640--641} (\bibinfo {year}
  {1947})}\BibitemShut {NoStop}%
\bibitem [{\citenamefont {Kienitz}\ \emph {et~al.}(2017)\citenamefont
  {Kienitz}, \citenamefont {D{\l}ugo{\l}\k{e}cki}, \citenamefont {Trippel},\
  and\ \citenamefont {K{\"u}pper}}]{Kienitz:JCP147:024304}%
  \BibitemOpen
  \bibfield  {author} {\bibinfo {author} {\bibfnamefont {J.~S.}\ \bibnamefont
  {Kienitz}}, \bibinfo {author} {\bibfnamefont {K.}~\bibnamefont
  {D{\l}ugo{\l}\k{e}cki}}, \bibinfo {author} {\bibfnamefont {S.}~\bibnamefont
  {Trippel}}, \ and\ \bibinfo {author} {\bibfnamefont {J.}~\bibnamefont
  {K{\"u}pper}},\ }\bibfield  {title} {\enquote {\bibinfo {title} {Improved
  spatial separation of neutral molecules},}\ }\href {\doibase
  10.1063/1.4991479} {\bibfield  {journal} {\bibinfo  {journal} {J.\ Chem.\
  Phys.}\ }\textbf {\bibinfo {volume} {147}},\ \bibinfo {pages} {024304}
  (\bibinfo {year} {2017})},\ \Eprint {http://arxiv.org/abs/1704.08912}
  {arXiv:1704.08912 [physics]} \BibitemShut {NoStop}%
\bibitem [{\citenamefont {Chang}\ \emph {et~al.}(2015)\citenamefont {Chang},
  \citenamefont {Horke}, \citenamefont {Trippel},\ and\ \citenamefont
  {K{\"u}pper}}]{Chang:IRPC34:557}%
  \BibitemOpen
  \bibfield  {author} {\bibinfo {author} {\bibfnamefont {Y.-P.}\ \bibnamefont
  {Chang}}, \bibinfo {author} {\bibfnamefont {D.~A.}\ \bibnamefont {Horke}},
  \bibinfo {author} {\bibfnamefont {S.}~\bibnamefont {Trippel}}, \ and\
  \bibinfo {author} {\bibfnamefont {J.}~\bibnamefont {K{\"u}pper}},\ }\bibfield
   {title} {\enquote {\bibinfo {title} {Spatially-controlled complex molecules
  and their applications},}\ }\href {\doibase 10.1080/0144235X.2015.1077838}
  {\bibfield  {journal} {\bibinfo  {journal} {Int.\ Rev.\ Phys.\ Chem.}\
  }\textbf {\bibinfo {volume} {34}},\ \bibinfo {pages} {557--590} (\bibinfo
  {year} {2015})},\ \Eprint {http://arxiv.org/abs/1505.05632} {arXiv:1505.05632
  [physics]} \BibitemShut {NoStop}%
\bibitem [{\citenamefont {Teschmit}, \citenamefont {Horke},\ and\ \citenamefont
  {Küpper}(2018)}]{Teschmit:ACIE57:13775}%
  \BibitemOpen
  \bibfield  {author} {\bibinfo {author} {\bibfnamefont {N.}~\bibnamefont
  {Teschmit}}, \bibinfo {author} {\bibfnamefont {D.~A.}\ \bibnamefont {Horke}},
  \ and\ \bibinfo {author} {\bibfnamefont {J.}~\bibnamefont {Küpper}},\
  }\bibfield  {title} {\enquote {\bibinfo {title} {Spatially separating the
  conformers of the dipeptide {Ac-Phe-Cys-NH$_2$}},}\ }\href {\doibase
  10.1002/anie.201807646} {\bibfield  {journal} {\bibinfo  {journal} {Angew.\
  Chem.\ Int.\ Ed.}\ }\textbf {\bibinfo {volume} {57}},\ \bibinfo {pages}
  {13775--13779} (\bibinfo {year} {2018})},\ \Eprint
  {http://arxiv.org/abs/1805.12396} {arXiv:1805.12396 [physics]} \BibitemShut
  {NoStop}%
\bibitem [{\citenamefont {Trippel}\ \emph {et~al.}(2018)\citenamefont
  {Trippel}, \citenamefont {Johny}, \citenamefont {Kierspel}, \citenamefont
  {Onvlee}, \citenamefont {Bieker}, \citenamefont {Ye}, \citenamefont
  {Mullins}, \citenamefont {Gumprecht}, \citenamefont {D{\l}ugo{\l}\k{e}cki},\
  and\ \citenamefont {K{\"u}pper}}]{Trippel:RSI89:096110}%
  \BibitemOpen
  \bibfield  {author} {\bibinfo {author} {\bibfnamefont {S.}~\bibnamefont
  {Trippel}}, \bibinfo {author} {\bibfnamefont {M.}~\bibnamefont {Johny}},
  \bibinfo {author} {\bibfnamefont {T.}~\bibnamefont {Kierspel}}, \bibinfo
  {author} {\bibfnamefont {J.}~\bibnamefont {Onvlee}}, \bibinfo {author}
  {\bibfnamefont {H.}~\bibnamefont {Bieker}}, \bibinfo {author} {\bibfnamefont
  {H.}~\bibnamefont {Ye}}, \bibinfo {author} {\bibfnamefont {T.}~\bibnamefont
  {Mullins}}, \bibinfo {author} {\bibfnamefont {L.}~\bibnamefont {Gumprecht}},
  \bibinfo {author} {\bibfnamefont {K.}~\bibnamefont {D{\l}ugo{\l}\k{e}cki}}, \
  and\ \bibinfo {author} {\bibfnamefont {J.}~\bibnamefont {K{\"u}pper}},\
  }\bibfield  {title} {\enquote {\bibinfo {title} {Knife edge skimming for
  improved separation of molecular species by the deflector},}\ }\href
  {\doibase 10.1063/1.5026145} {\bibfield  {journal} {\bibinfo  {journal}
  {Rev.\ Sci.\ Instrum.}\ }\textbf {\bibinfo {volume} {89}},\ \bibinfo {pages}
  {096110} (\bibinfo {year} {2018})},\ \Eprint
  {http://arxiv.org/abs/1802.04053} {arXiv:1802.04053 [physics]} \BibitemShut
  {NoStop}%
\bibitem [{\citenamefont {Nielsen}\ \emph {et~al.}(2011)\citenamefont
  {Nielsen}, \citenamefont {Simesen}, \citenamefont {Bisgaard}, \citenamefont
  {Stapelfeldt}, \citenamefont {Filsinger}, \citenamefont {Friedrich},
  \citenamefont {Meijer},\ and\ \citenamefont
  {K{\"u}pper}}]{Nielsen:PCCP13:18971}%
  \BibitemOpen
  \bibfield  {author} {\bibinfo {author} {\bibfnamefont {J.~H.}\ \bibnamefont
  {Nielsen}}, \bibinfo {author} {\bibfnamefont {P.}~\bibnamefont {Simesen}},
  \bibinfo {author} {\bibfnamefont {C.~Z.}\ \bibnamefont {Bisgaard}}, \bibinfo
  {author} {\bibfnamefont {H.}~\bibnamefont {Stapelfeldt}}, \bibinfo {author}
  {\bibfnamefont {F.}~\bibnamefont {Filsinger}}, \bibinfo {author}
  {\bibfnamefont {B.}~\bibnamefont {Friedrich}}, \bibinfo {author}
  {\bibfnamefont {G.}~\bibnamefont {Meijer}}, \ and\ \bibinfo {author}
  {\bibfnamefont {J.}~\bibnamefont {K{\"u}pper}},\ }\bibfield  {title}
  {\enquote {\bibinfo {title} {Stark-selected beam of ground-state {OCS}
  molecules characterized by revivals of impulsive alignment},}\ }\href
  {\doibase 10.1039/c1cp21143a} {\bibfield  {journal} {\bibinfo  {journal}
  {Phys.\ Chem.\ Chem.\ Phys.}\ }\textbf {\bibinfo {volume} {13}},\ \bibinfo
  {pages} {18971--18975} (\bibinfo {year} {2011})},\ \Eprint
  {http://arxiv.org/abs/1105.2413} {arXiv:1105.2413 [physics]} \BibitemShut
  {NoStop}%
\bibitem [{\citenamefont {Karamatskos}\ \emph {et~al.}(2018)\citenamefont
  {Karamatskos}, \citenamefont {Raabe}, \citenamefont {Mullins}, \citenamefont
  {Trabattoni}, \citenamefont {Stammer}, \citenamefont {Goldsztejn},
  \citenamefont {Johansen}, \citenamefont {Długołęcki}, \citenamefont
  {Stapelfeldt}, \citenamefont {Vrakking}, \citenamefont {Trippel},
  \citenamefont {Rouzée},\ and\ \citenamefont
  {Küpper}}]{Karamatskos:arXiv1807:01034}%
  \BibitemOpen
  \bibfield  {author} {\bibinfo {author} {\bibfnamefont {E.~T.}\ \bibnamefont
  {Karamatskos}}, \bibinfo {author} {\bibfnamefont {S.}~\bibnamefont {Raabe}},
  \bibinfo {author} {\bibfnamefont {T.}~\bibnamefont {Mullins}}, \bibinfo
  {author} {\bibfnamefont {A.}~\bibnamefont {Trabattoni}}, \bibinfo {author}
  {\bibfnamefont {P.}~\bibnamefont {Stammer}}, \bibinfo {author} {\bibfnamefont
  {G.}~\bibnamefont {Goldsztejn}}, \bibinfo {author} {\bibfnamefont {R.~R.}\
  \bibnamefont {Johansen}}, \bibinfo {author} {\bibfnamefont {K.}~\bibnamefont
  {Długołęcki}}, \bibinfo {author} {\bibfnamefont {H.}~\bibnamefont
  {Stapelfeldt}}, \bibinfo {author} {\bibfnamefont {M.~J.~J.}\ \bibnamefont
  {Vrakking}}, \bibinfo {author} {\bibfnamefont {S.}~\bibnamefont {Trippel}},
  \bibinfo {author} {\bibfnamefont {A.}~\bibnamefont {Rouzée}}, \ and\
  \bibinfo {author} {\bibfnamefont {J.}~\bibnamefont {Küpper}},\ }\bibfield
  {title} {\enquote {\bibinfo {title} {Molecular movie of ultrafast coherent
  rotational dynamics},}\ }\href {https://arxiv.org/pdf/1807.01034.pdf} {\
  (\bibinfo {year} {2018})},\ \Eprint {http://arxiv.org/abs/1807.01034}
  {arXiv:1807.01034} \BibitemShut {NoStop}%
\bibitem [{\citenamefont {Dribinski}\ \emph {et~al.}(2002)\citenamefont
  {Dribinski}, \citenamefont {Ossadtchi}, \citenamefont {Mandelshtam},\ and\
  \citenamefont {Reisler}}]{Dribinski:RSI73:2634}%
  \BibitemOpen
  \bibfield  {author} {\bibinfo {author} {\bibfnamefont {V.}~\bibnamefont
  {Dribinski}}, \bibinfo {author} {\bibfnamefont {A.}~\bibnamefont
  {Ossadtchi}}, \bibinfo {author} {\bibfnamefont {V.~A.}\ \bibnamefont
  {Mandelshtam}}, \ and\ \bibinfo {author} {\bibfnamefont {H.}~\bibnamefont
  {Reisler}},\ }\bibfield  {title} {\enquote {\bibinfo {title} {{Reconstruction
  of {A}bel-transformable images: {T}he {G}aussian basis-set expansion {A}bel
  transform method}},}\ }\href {\doibase 10.1063/1.1482156} {\bibfield
  {journal} {\bibinfo  {journal} {Rev.\ Sci.\ Instrum.}\ }\textbf {\bibinfo
  {volume} {73}},\ \bibinfo {pages} {2634} (\bibinfo {year}
  {2002})}\BibitemShut {NoStop}%
\bibitem [{\citenamefont {Chen}\ \emph {et~al.}(2009)\citenamefont {Chen},
  \citenamefont {Le}, \citenamefont {Morishita},\ and\ \citenamefont
  {Lin}}]{Chen:PRA79:033409}%
  \BibitemOpen
  \bibfield  {author} {\bibinfo {author} {\bibfnamefont {Z.}~\bibnamefont
  {Chen}}, \bibinfo {author} {\bibfnamefont {A.-T.}\ \bibnamefont {Le}},
  \bibinfo {author} {\bibfnamefont {T.}~\bibnamefont {Morishita}}, \ and\
  \bibinfo {author} {\bibfnamefont {C.~D.}\ \bibnamefont {Lin}},\ }\bibfield
  {title} {\enquote {\bibinfo {title} {Quantitative rescattering theory for
  laser-induced high-energy plateau photoelectron spectra},}\ }\href {\doibase
  10.1103/PhysRevA.79.033409} {\bibfield  {journal} {\bibinfo  {journal}
  {Phys.\ Rev.\ A}\ }\textbf {\bibinfo {volume} {79}},\ \bibinfo {pages}
  {033409} (\bibinfo {year} {2009})}\BibitemShut {NoStop}%
\bibitem [{\citenamefont {Xu}\ \emph {et~al.}(2010)\citenamefont {Xu},
  \citenamefont {Chen}, \citenamefont {Le},\ and\ \citenamefont
  {Lin}}]{Xu:PRA82:033403}%
  \BibitemOpen
  \bibfield  {author} {\bibinfo {author} {\bibfnamefont {J.}~\bibnamefont
  {Xu}}, \bibinfo {author} {\bibfnamefont {Z.}~\bibnamefont {Chen}}, \bibinfo
  {author} {\bibfnamefont {A.-T.}\ \bibnamefont {Le}}, \ and\ \bibinfo {author}
  {\bibfnamefont {C.~D.}\ \bibnamefont {Lin}},\ }\bibfield  {title} {\enquote
  {\bibinfo {title} {Self-imaging of molecules from diffraction spectra by
  laser-induced rescattering electrons},}\ }\href {\doibase
  10.1103/PhysRevA.82.033403} {\bibfield  {journal} {\bibinfo  {journal} {Phys.
  Rev. A}\ }\textbf {\bibinfo {volume} {82}},\ \bibinfo {pages} {033403}
  (\bibinfo {year} {2010})}\BibitemShut {NoStop}%
\bibitem [{\citenamefont {Fon}\ \emph {et~al.}(1983)\citenamefont {Fon},
  \citenamefont {Berrington}, \citenamefont {Burke},\ and\ \citenamefont
  {Hibbert}}]{Fon:JPB16:307}%
  \BibitemOpen
  \bibfield  {author} {\bibinfo {author} {\bibfnamefont {W.~C.}\ \bibnamefont
  {Fon}}, \bibinfo {author} {\bibfnamefont {K.~A.}\ \bibnamefont {Berrington}},
  \bibinfo {author} {\bibfnamefont {P.~G.}\ \bibnamefont {Burke}}, \ and\
  \bibinfo {author} {\bibfnamefont {A.}~\bibnamefont {Hibbert}},\ }\bibfield
  {title} {\enquote {\bibinfo {title} {The elastic scattering of electrons from
  inert gases. {III}. {A}rgon},}\ }\href {\doibase
  {10.1088/0022-3700/16/2/018}} {\bibfield  {journal} {\bibinfo  {journal} {J.\
  Phys.\ B}\ }\textbf {\bibinfo {volume} {16}},\ \bibinfo {pages} {307}
  (\bibinfo {year} {1983})}\BibitemShut {NoStop}%
\bibitem [{\citenamefont {Fon}, \citenamefont {Berrington},\ and\ \citenamefont
  {Hibbert}(1984)}]{Fon:JPB17:3279}%
  \BibitemOpen
  \bibfield  {author} {\bibinfo {author} {\bibfnamefont {W.~C.}\ \bibnamefont
  {Fon}}, \bibinfo {author} {\bibfnamefont {K.~A.}\ \bibnamefont {Berrington}},
  \ and\ \bibinfo {author} {\bibfnamefont {A.}~\bibnamefont {Hibbert}},\
  }\bibfield  {title} {\enquote {\bibinfo {title} {The elastic scattering of
  electrons from inert gases. {IV}. {K}rypton},}\ }\href {\doibase
  10.1088/0022-3700/17/16/011} {\bibfield  {journal} {\bibinfo  {journal} {J.\
  Phys.\ B}\ }\textbf {\bibinfo {volume} {17}},\ \bibinfo {pages} {3279--3294}
  (\bibinfo {year} {1984})}\BibitemShut {NoStop}%
\bibitem [{\citenamefont {Murai}\ \emph {et~al.}(2013)\citenamefont {Murai},
  \citenamefont {Ishijima}, \citenamefont {Mitsumura}, \citenamefont
  {Sakamoto}, \citenamefont {Kato}, \citenamefont {Hoshino}, \citenamefont
  {Blanco}, \citenamefont {Garci­a}, \citenamefont {Limao-Vieira},
  \citenamefont {Brunger}, \citenamefont {Buckman},\ and\ \citenamefont
  {Tanaka}}]{Mirai:JCP5:054302}%
  \BibitemOpen
  \bibfield  {author} {\bibinfo {author} {\bibfnamefont {H.}~\bibnamefont
  {Murai}}, \bibinfo {author} {\bibfnamefont {Y.}~\bibnamefont {Ishijima}},
  \bibinfo {author} {\bibfnamefont {T.}~\bibnamefont {Mitsumura}}, \bibinfo
  {author} {\bibfnamefont {Y.}~\bibnamefont {Sakamoto}}, \bibinfo {author}
  {\bibfnamefont {H.}~\bibnamefont {Kato}}, \bibinfo {author} {\bibfnamefont
  {M.}~\bibnamefont {Hoshino}}, \bibinfo {author} {\bibfnamefont
  {F.}~\bibnamefont {Blanco}}, \bibinfo {author} {\bibfnamefont
  {G.}~\bibnamefont {Garci­a}}, \bibinfo {author} {\bibfnamefont
  {P.}~\bibnamefont {Limao-Vieira}}, \bibinfo {author} {\bibfnamefont {M.~J.}\
  \bibnamefont {Brunger}}, \bibinfo {author} {\bibfnamefont {S.~J.}\
  \bibnamefont {Buckman}}, \ and\ \bibinfo {author} {\bibfnamefont
  {H.}~\bibnamefont {Tanaka}},\ }\bibfield  {title} {\enquote {\bibinfo {title}
  {A comprehensive and comparative study of elastic electron scattering from
  {OCS} and {CS}$_2$ in the energy region from 1.2 to 200 e{V}},}\ }\href
  {\doibase 10.1063/1.4788666} {\bibfield  {journal} {\bibinfo  {journal} {J.\
  Chem.\ Phys.}\ }\textbf {\bibinfo {volume} {138}},\ \bibinfo {pages} {054302}
  (\bibinfo {year} {2013})}\BibitemShut {NoStop}%
\bibitem [{\citenamefont {Michelin}\ \emph {et~al.}(2000)\citenamefont
  {Michelin}, \citenamefont {Kroin}, \citenamefont {Iga}, \citenamefont
  {Homem}, \citenamefont {Miglio},\ and\ \citenamefont
  {Lee}}]{Michelin:JPB33:3293}%
  \BibitemOpen
  \bibfield  {author} {\bibinfo {author} {\bibfnamefont {S.~E.}\ \bibnamefont
  {Michelin}}, \bibinfo {author} {\bibfnamefont {T.}~\bibnamefont {Kroin}},
  \bibinfo {author} {\bibfnamefont {I.}~\bibnamefont {Iga}}, \bibinfo {author}
  {\bibfnamefont {M.~G.~P.}\ \bibnamefont {Homem}}, \bibinfo {author}
  {\bibfnamefont {H.~S.}\ \bibnamefont {Miglio}}, \ and\ \bibinfo {author}
  {\bibfnamefont {M.~T.}\ \bibnamefont {Lee}},\ }\bibfield  {title} {\enquote
  {\bibinfo {title} {Elastic and total cross sections for electron-carbonyl
  sulfide collisions},}\ }\href {http://stacks.iop.org/0953-4075/33/i=17/a=310}
  {\bibfield  {journal} {\bibinfo  {journal} {J.\ Phys.\ B}\ }\textbf {\bibinfo
  {volume} {33}},\ \bibinfo {pages} {3293} (\bibinfo {year}
  {2000})}\BibitemShut {NoStop}%
\bibitem [{\citenamefont {Schell}\ \emph {et~al.}(2018)\citenamefont {Schell},
  \citenamefont {Bredtmann}, \citenamefont {Schulz}, \citenamefont
  {Patchkovskii}, \citenamefont {Vrakking},\ and\ \citenamefont
  {Mikosch}}]{Schell:SciAdv4:eaap8148}%
  \BibitemOpen
  \bibfield  {author} {\bibinfo {author} {\bibfnamefont {F.}~\bibnamefont
  {Schell}}, \bibinfo {author} {\bibfnamefont {T.}~\bibnamefont {Bredtmann}},
  \bibinfo {author} {\bibfnamefont {C.-P.}\ \bibnamefont {Schulz}}, \bibinfo
  {author} {\bibfnamefont {S.}~\bibnamefont {Patchkovskii}}, \bibinfo {author}
  {\bibfnamefont {M.~J.~J.}\ \bibnamefont {Vrakking}}, \ and\ \bibinfo {author}
  {\bibfnamefont {J.}~\bibnamefont {Mikosch}},\ }\bibfield  {title} {\enquote
  {\bibinfo {title} {Molecular orbital imprint in laser-driven electron
  recollision},}\ }\href {\doibase 10.1126/sciadv.aap8148} {\bibfield
  {journal} {\bibinfo  {journal} {Science Advances}\ }\textbf {\bibinfo
  {volume} {4}},\ \bibinfo {pages} {eaap8148} (\bibinfo {year}
  {2018})}\BibitemShut {NoStop}%
\bibitem [{\citenamefont {Kre\ifmmode \check{c}\else
  \v{c}\fi{}ini\ifmmode~\acute{c}\else \'{c}\fi{}}\ \emph
  {et~al.}(2018)\citenamefont {Kre\ifmmode \check{c}\else
  \v{c}\fi{}ini\ifmmode~\acute{c}\else \'{c}\fi{}}, \citenamefont {Wopperer},
  \citenamefont {Frusteri}, \citenamefont {Brau\ss{}e}, \citenamefont
  {Brisset}, \citenamefont {De~Giovannini}, \citenamefont {Rubio},
  \citenamefont {Rouz\'ee},\ and\ \citenamefont
  {Vrakking}}]{Krecinic:PRA98:041401}%
  \BibitemOpen
  \bibfield  {author} {\bibinfo {author} {\bibfnamefont {F.}~\bibnamefont
  {Kre\ifmmode \check{c}\else \v{c}\fi{}ini\ifmmode~\acute{c}\else
  \'{c}\fi{}}}, \bibinfo {author} {\bibfnamefont {P.}~\bibnamefont {Wopperer}},
  \bibinfo {author} {\bibfnamefont {B.}~\bibnamefont {Frusteri}}, \bibinfo
  {author} {\bibfnamefont {F.}~\bibnamefont {Brau\ss{}e}}, \bibinfo {author}
  {\bibfnamefont {J.-G.}\ \bibnamefont {Brisset}}, \bibinfo {author}
  {\bibfnamefont {U.}~\bibnamefont {De~Giovannini}}, \bibinfo {author}
  {\bibfnamefont {A.}~\bibnamefont {Rubio}}, \bibinfo {author} {\bibfnamefont
  {A.}~\bibnamefont {Rouz\'ee}}, \ and\ \bibinfo {author} {\bibfnamefont
  {M.~J.~J.}\ \bibnamefont {Vrakking}},\ }\bibfield  {title} {\enquote
  {\bibinfo {title} {Multiple-orbital effects in laser-induced electron
  diffraction of aligned molecules},}\ }\href {\doibase
  10.1103/PhysRevA.98.041401} {\bibfield  {journal} {\bibinfo  {journal}
  {Phys.\ Rev.\ A}\ }\textbf {\bibinfo {volume} {98}},\ \bibinfo {pages}
  {041401} (\bibinfo {year} {2018})}\BibitemShut {NoStop}%
\bibitem [{\citenamefont {Trabattoni}\ \emph {et~al.}(2018)\citenamefont
  {Trabattoni}, \citenamefont {Trippel}, \citenamefont {Giovannini},
  \citenamefont {Olivieri}, \citenamefont {Wiese}, \citenamefont {Mullins},
  \citenamefont {Onvlee}, \citenamefont {Son}, \citenamefont
  {Biagio~Frusteri},\ and\ \citenamefont
  {K{\"u}pper}}]{Trabattoni:cutoff:inprep}%
  \BibitemOpen
  \bibfield  {author} {\bibinfo {author} {\bibfnamefont {A.}~\bibnamefont
  {Trabattoni}}, \bibinfo {author} {\bibfnamefont {S.}~\bibnamefont {Trippel}},
  \bibinfo {author} {\bibfnamefont {U.~D.}\ \bibnamefont {Giovannini}},
  \bibinfo {author} {\bibfnamefont {J.~F.}\ \bibnamefont {Olivieri}}, \bibinfo
  {author} {\bibfnamefont {J.}~\bibnamefont {Wiese}}, \bibinfo {author}
  {\bibfnamefont {T.}~\bibnamefont {Mullins}}, \bibinfo {author} {\bibfnamefont
  {J.}~\bibnamefont {Onvlee}}, \bibinfo {author} {\bibfnamefont {S.-K.}\
  \bibnamefont {Son}}, \bibinfo {author} {\bibfnamefont {A.~R.}\ \bibnamefont
  {Biagio~Frusteri}}, \ and\ \bibinfo {author} {\bibfnamefont {J.}~\bibnamefont
  {K{\"u}pper}},\ }\bibfield  {title} {\enquote {\bibinfo {title} {Setting the
  clock of photoelectron emission through molecular alignment},}\ }\href
  {https://arxiv.org/abs/1802.06622} {\  (\bibinfo {year} {2018})},\ \Eprint
  {http://arxiv.org/abs/1802.06622} {arXiv:1802.06622 [physics]} \BibitemShut
  {NoStop}%
\bibitem [{\citenamefont {Holmegaard}\ \emph {et~al.}(2010)\citenamefont
  {Holmegaard}, \citenamefont {Hansen}, \citenamefont {Kalhoj}, \citenamefont
  {Kragh}, \citenamefont {Stapelfeldt}, \citenamefont {Filsinger},
  \citenamefont {K{\"u}pper}, \citenamefont {Meijer}, \citenamefont
  {Dimitrovski}, \citenamefont {Abu-samha}, \citenamefont {Martiny},\ and\
  \citenamefont {Madsen}}]{Holmegaard:NatPhys:2010}%
  \BibitemOpen
  \bibfield  {author} {\bibinfo {author} {\bibfnamefont {L.}~\bibnamefont
  {Holmegaard}}, \bibinfo {author} {\bibfnamefont {J.~L.}\ \bibnamefont
  {Hansen}}, \bibinfo {author} {\bibfnamefont {L.}~\bibnamefont {Kalhoj}},
  \bibinfo {author} {\bibfnamefont {S.~L.}\ \bibnamefont {Kragh}}, \bibinfo
  {author} {\bibfnamefont {H.}~\bibnamefont {Stapelfeldt}}, \bibinfo {author}
  {\bibfnamefont {F.}~\bibnamefont {Filsinger}}, \bibinfo {author}
  {\bibfnamefont {J.}~\bibnamefont {K{\"u}pper}}, \bibinfo {author}
  {\bibfnamefont {G.}~\bibnamefont {Meijer}}, \bibinfo {author} {\bibfnamefont
  {D.}~\bibnamefont {Dimitrovski}}, \bibinfo {author} {\bibfnamefont
  {M.}~\bibnamefont {Abu-samha}}, \bibinfo {author} {\bibfnamefont {C.~P.~J.}\
  \bibnamefont {Martiny}}, \ and\ \bibinfo {author} {\bibfnamefont {L.~B.}\
  \bibnamefont {Madsen}},\ }\bibfield  {title} {\enquote {\bibinfo {title}
  {Photoelectron angular distributions from strong-field ionization of oriented
  molecules},}\ }\href {\doibase 10.1038/NPHYS1666} {\bibfield  {journal}
  {\bibinfo  {journal} {Nat. Phys.}\ }\textbf {\bibinfo {volume} {6}},\
  \bibinfo {pages} {428} (\bibinfo {year} {2010})}\BibitemShut {NoStop}%
\bibitem [{\citenamefont {Bilalbegovic}(2008)}]{Bilalbegovic:EPJD49:43}%
  \BibitemOpen
  \bibfield  {author} {\bibinfo {author} {\bibfnamefont {G.}~\bibnamefont
  {Bilalbegovic}},\ }\bibfield  {title} {\enquote {\bibinfo {title} {Carbonyl
  sulphide under strong laser field: time-dependent density functional
  theory},}\ }\href {\doibase 10.1140/epjd/e2008-00137-8} {\bibfield  {journal}
  {\bibinfo  {journal} {Eur.\ Phys.\ J.\ D}\ }\textbf {\bibinfo {volume}
  {49}},\ \bibinfo {pages} {43--49} (\bibinfo {year} {2008})}\BibitemShut
  {NoStop}%
\bibitem [{\citenamefont {Hargittai}\ and\ \citenamefont
  {Hargittai}(1988)}]{Hargittai:GED:1988}%
  \BibitemOpen
  \bibfield  {author} {\bibinfo {author} {\bibfnamefont {I.}~\bibnamefont
  {Hargittai}}\ and\ \bibinfo {author} {\bibfnamefont {M.}~\bibnamefont
  {Hargittai}},\ }\href
  {http://www.wiley-vch.de/publish/en/books/ISBN978-0-471-18689-2} {\emph
  {\bibinfo {title} {Stereochemical Applications of Gas-Phase Electron
  Diffraction}}}\ (\bibinfo  {publisher} {VCH Verlagsgesellschaft},\ \bibinfo
  {address} {Weinheim, Germany},\ \bibinfo {year} {1988})\BibitemShut {NoStop}%
\bibitem [{\citenamefont {Lin}\ \emph {et~al.}(2010)\citenamefont {Lin},
  \citenamefont {Le}, \citenamefont {Chen}, \citenamefont {Morishita},\ and\
  \citenamefont {Lucchese}}]{Lin:JPB43:122001}%
  \BibitemOpen
  \bibfield  {author} {\bibinfo {author} {\bibfnamefont {C.~D.}\ \bibnamefont
  {Lin}}, \bibinfo {author} {\bibfnamefont {A.-T.}\ \bibnamefont {Le}},
  \bibinfo {author} {\bibfnamefont {Z.}~\bibnamefont {Chen}}, \bibinfo {author}
  {\bibfnamefont {T.}~\bibnamefont {Morishita}}, \ and\ \bibinfo {author}
  {\bibfnamefont {R.}~\bibnamefont {Lucchese}},\ }\bibfield  {title} {\enquote
  {\bibinfo {title} {Strong-field rescattering physics - self-imaging of a
  molecule by its own electrons},}\ }\href {\doibase
  10.1088/0953-4075/43/12/122001} {\bibfield  {journal} {\bibinfo  {journal}
  {J.\ Phys.\ B}\ }\textbf {\bibinfo {volume} {43}},\ \bibinfo {pages} {122001}
  (\bibinfo {year} {2010})}\BibitemShut {NoStop}%
\end{thebibliography}%
\onecolumngrid%
\end{document}